\documentclass[graybox]{svmult}
\usepackage{graphicx}
\usepackage{multicol}
\usepackage[bottom]{footmisc}
\usepackage{comment}
\usepackage{textcomp}
\usepackage{newtxtext} 
\usepackage{newtxmath}
\newcommand{\mean}[1]{\ensuremath{\left\langle #1 \right\rangle}}
\newcommand{\ket}[1]{\ensuremath{| #1 \rangle}}
\usepackage[square,numbers]{natbib}
\makeatletter \renewcommand\@biblabel[1]{#1.} \makeatother
\usepackage{url} 
\makeindex
\begin{document}
\title*{Attractive Multidimensional Condensates--Experiments}
\titlerunning{Attractive Multidimensional Condensates--Experiments}
\author{
Hikaru Tamura
\and
Chen-Lung Hung
}
\institute{
Hikaru Tamura
\at
Institute for Molecular Science, Okazaki, Aichi 444-8585, Japan
\email{tamurah@ims.ac.jp}
\and
Chen-Lung Hung
\at
Purdue University, West Lafayette, Indiana 47907, USA
\email{clhung@purdue.edu}
}
\maketitle
\abstract
{
Experiments on attractive Bose-Einstein condensates (BECs) have unlocked many intriguing out-of-equilibrium dynamics through the interplay between matter-wave dispersion and nonlinear attractive interaction. Competition between these effects leads to fascinating phenomena such as wave collapse, modulational instability, and formation of multidimensional bright solitons. This chapter reviews experimental studies on attractive condensates, with a primary focus on alkali atoms featuring two-body contact interactions. We review recent experimental advances in optical trapping and interaction control techniques, which have enabled new studies on attractive condensates in three and also in lower dimensions. Specifically, we discuss pioneering and recent experimental observations on the dynamics and stability of attractive BECs, including the formation of bright solitons, their collisions, and excitations in quasi-one-dimensional traps. Recent observations of the elusive two-dimensional Townes solitons and vortex solitons are also discussed in this Chapter. We then highlight an experimental technique revealing the nonclassical signatures of modulational instability in an attractive condensate.
}

\section{Introduction}
Attractive Bose-Einstein condensates (BECs), characterized by a negative $s$-wave scattering length $a_{s}$ between the atoms, serve not only as a universal testbed for classical nonlinear physics with self-focusing nonlinearity, but also as a bridge between macroscopic phenomena and quantum many-body dynamics, where quantum correlations and fluctuations play critical roles. Unlike repulsively interacting BECs that can be stably produced with well-defined energies, attractive Bose gases tend to collapse to higher densities to lower the interaction energy. In an attractive condensate, density perturbations at length scales larger than the condensate's healing length can also self-amplify owing to modulational instability. Consequently, both global density variation and local density fluctuations can have profound impact on the stability and collapse dynamics of attractive condensates. The ability in quantum gas experiments to control the atomic interaction, confining potential, and dimensionality gives rise to unprecedented opportunities to explore the rich dynamics at attractive interactions, which is the focus of this chapter. 

The behavior of attractive condensates is strongly influenced by the dimensionality. Here, we consider a $D$-dimensional condensate wavefunction $\psi(\mathbf{r})$ governed by the Gross–Pitaevskii equation (GPE) in the absence of trap confinement. Following the argument of Derrick's theorem~\cite{derrick1964comments}, the stability for the condensate can be seen from its energy functional,
\begin{eqnarray}
E&=&\int \left[  \frac{\hbar^{2}}{2m} \left| \nabla \psi \right|^{2} + \frac{g}{2} \left| \psi \right|^{4}\right]d^{D} r, \label{eq:energyfunc}
\end{eqnarray}
where $m$ is the atomic mass, $\hbar=h/(2\pi) $ is the reduced Planck constant, and $g=4\pi \hbar^{2} a_{s}/m<0$ is the interaction parameter for the mean-field attraction. From a scaling analysis assuming a wavepacket of size $L$ and norm (atom number) $N=\int \left| \psi \right|^{2} d^{D} r$, the kinetic energy and the interaction energy scale as $L^{-2}$ and $L^{-D}$, respectively. For a stable solution to exist, the energy functional should exhibit a local minimum satisfying $\partial E / \partial L = 0$ and $\partial^{2}E/\partial L^{2}>0$. There is no stable solution in the three-dimensional (3D) case, since $\partial^{2} E/\partial L^{2}<0$ for any extremum. This suggests any 3D condensate must be dynamically unstable and collapses below a critical size. In the one-dimensional (1D) case, a stable minimum can be found, indicating that the attraction can be balanced by the kinetic energy and the wavefunction does not collapse. A stable 1D solution is commonly referred to as a \emph{bright soliton}. The two-dimensional (2D) GPE, on the other hand, lacks a stable solution as both energy terms scale as $L^{-2}$, which also gives rise to scale invariance\footnote{In 2D GPE, rescaling a wavefunction $\psi\rightarrow \psi/\Lambda$ and spacetime coordinates $(\mathbf{r},t)\rightarrow (\Lambda\mathbf{r}, \Lambda^2t)$ by an arbitrary scale factor $\Lambda$ results in identical, scale-invariant dynamics.}. A wavefunction may either collapse or disperse, depending solely on the norm, not on its physical size. However, a wavefunction may become stationary at a specific waveform and norm, when the two energy terms precisely cancel each other. This solution is commonly referred to as the \emph{Townes soliton}.

Since the realization of atomic BECs, experimental techniques have advanced to enable studies of multidimensional attractive condensates. Besides early application of magnetic field trapping for condensate production, \emph{optical dipole trapping} rapidly became one of the most important experimental tools~\cite{grimm2000optical}. It not only made experiments on magnetically non-trappable atomic states possible, but also enabled flexible trap control and reduction of dimensionality. In an optical dipole trap, strong and far-off-resonant light induces an electric dipole potential proportional to the local light intensity. Red-detuned light attracts atoms to the local intensity maxima while blue-detuned light repels them to the local minima. 
Using tightly focused beams or interference light patterns featuring steep intensity variations in micrometer length scales, cigar-shaped quasi-1D traps or pancake-like quasi-2D traps can be created. A 3D BEC can be compressed to form a low-dimensional condensate, when the transverse trap length scale is made smaller than the condensate's healing length. 

\emph{Magnetic Feshbach resonance} provides a powerful interaction tuning knob for creating attractive condensates~\cite{Chin2010FeshbachGases}. A Feshbach resonance occurs when two colliding atoms become energetically resonant to a diatomic molecular state of a different spin configuration. An inherent spin-exchange coupling mixes the atomic state with the molecular bound state, thus modifying the collisional phase shift and effectively changing the s-wave scattering length. Via a magnetic Feshbach resonance, the scattering length $a_s(B)$ can be dynamically controlled as a function of an external magnetic field $B$, which adjusts the relative Zeeman energy shift between the atoms and the resonant bound state to modify the scattering phase shift. Broader Feshbach resonances are typically utilized in experiments, as the scattering length can be smoothly tuned from positive to negative values without approaching too close to the resonance where inelastic collisions mediated by the short-lived molecular state become very severe. A typical experiment starts by producing a stable BEC at a positive interaction $a_s(B)>0$. The magnetic field is then ramped or quenched to a new value $B'$ to arrive at a negative $a_s(B')<0$, converting the BEC to an attractive condensate. To-date, nearly all experiments studying attractive condensates have utilized magnetic Feshbach resonances for interaction tuning. For alkali atoms, these include, but not limited to, $^7$Li~\cite{strecker2002formation,khaykovich2002formation,chen2019dynamical}, $^{39}$K~\cite{lepoutre2016production,eigen2016observation}, $^{85}$Rb~\cite{roberts2001controlled,marchant2013controlled,mcdonald2014bright}, $^{87}$Rb~\cite{compton2012dynamically}, and $^{133}$Cs~\cite{di2019excitation,mevznarvsivc2019cesium,Chen2020ObservationGases,huang2024two}.

This chapter reviews experimental studies on attractive condensates primarily formed by single-component alkali atoms, with a stronger emphasis on select recent observations. In Sect.~\ref{sec:box}, we briefly review the experimental creation of optical boxes, which has enabled new studies of multidimensional attractive condensates. In Sect.~\ref{sec:3d}, we discuss both pioneering and recent experimental observations on the stability and collapse dynamics of 3D BECs. Section~\ref{sec:1d} focuses on bright solitons in quasi-1D traps, where we discuss their formation mechanism, soliton collisions, and the excited soliton breathers. In Sect.~\ref{sec:MI}, modulational instability and the resulting wave collapse and soliton formation dynamics are discussed. Section~\ref{sec:2d} covers the observations of Townes solitons and vortex solitons in quasi-2D traps. In Sect.~\ref{sec:quantumMI}, we provide a detailed discussion of an experiment revealing the quantum dynamics of modulational instability and the hidden nonclassical signatures in condensate collapse dynamics. We conclude with an outlook discussion in Sect.~\ref{sec:outlook}. 

\section{Attractive condensates in optical boxes}
\label{sec:box}
\begin{figure}[b]
\sidecaption[c]
\includegraphics[scale=.45]{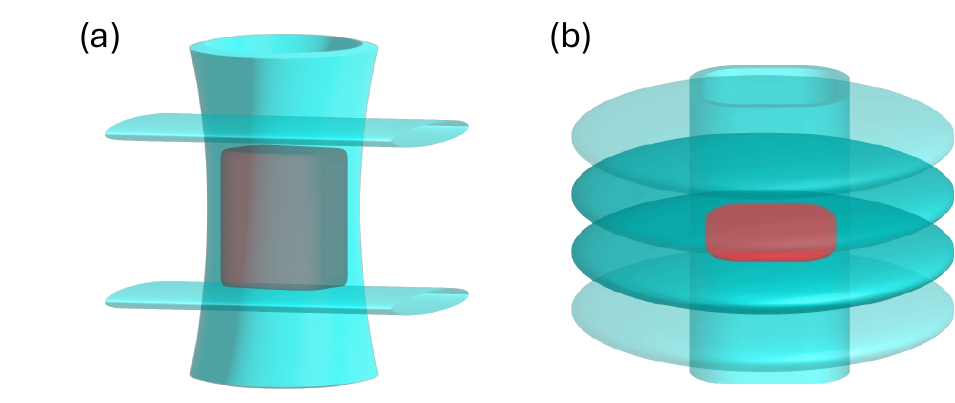}
\caption{Schematics of (a) 3D and (b) 2D optical boxes.}
\label{fig:box}     
\end{figure}

Optical box traps~\cite{navon2021quantum} present new ways to create and control attractive multidimensional condensates. The first optical box was developed by the Cambridge group~\cite{gaunt2013bose} using a blue-detuned, cylindrical hollow beam intersecting with two parallel light sheets, as illustrated in Fig.~\ref{fig:box}(a). These light beams exerted strong repulsive force on the atoms, forming `box walls' to confine a BEC in the enclosed dark region. At an initial repulsive atomic interaction $a_s>0$, the BEC formed with uniform density within the box. When the atomic interaction was tuned to attractive $a_s<0$, an initially uniform 3D BEC would display distinctive collapse dynamics different from those of nonuniform 3D gases in harmonic traps, discussed in Sect.~\ref{sec:3d}. 

Optical boxes were subsequently adapted to study two-dimensional attractive condensates~\cite{Chen2020ObservationGases,bakkali2021realization}. In these experiments, a pure BEC was first compressed into a quasi-2D gas. The compression can be achieved by ramping up two closely spaced repulsive barriers formed, for example, by a Hermite-Gauss beam~\cite{chomaz2015emergence}. Another way to impose strong confinement was through loading atoms into a single nodal plane of a blue-detuned 1D optical lattice with periodicity of a few microns~\cite{Chen2020ObservationGases,mitra2016phase,ville2017loading,hueck2018two}, which was formed by interfering two laser beams at a small angle; see Fig.~\ref{fig:box}(b). In the optical lattice, light intensity was minimized around a node, allowing atoms to move freely within the nodal plane while keeping the out-of-plane atomic motion in the harmonic ground state. To prevent gravity from tilting the potential, the lattice orientation was aligned vertically. Box walls were superimposed onto the 2D plane using a hollow light pattern, which was arbitrarily shaped by a digital micromirror device (DMD)~\cite{ville2017loading}, or by a spatial light modulator (SLM)~\cite{gaunt2013bose}, and projected through an objective lens with high optical resolution ($\lesssim 1~\mu\mathrm{m}$). Typical experiments were carried out in the weakly attractive regime, $|a_s|\ll l_z$, where $l_z$ is the vertical harmonic oscillator length. This experimental setting could be modeled by a standard 2D GPE with a cubic nonlinear interaction term, where the effective 2D interaction parameter $g_\mathrm{2D} = \sqrt{8\pi}a_s/l_z$ was obtained after integrating out the vertical atomic wave function.

As an extension of the 2D box technique, a 2D condensate can be loaded into an elongated box of narrow transverse width smaller than the healing length, effectively forming a quasi-1D gas of uniform density along the long axis of the box. Using a DMD~\cite{rabec2024bloch,tamura2025observation} or an SLM, multiple parallel 1D gases can be prepared in a single experiment. By removing the box walls while keeping the vertical lattice confinement, the 1D gases quickly expand transversely and interfere. This opens a way to probe phase coherence in attractive condensates through matter-wave interferometry. 

Besides collapse dynamics observed in three dimensions (Sect.~\ref{sec:collapse}), the adoption of optical boxes quickly led to observations of intriguing collapse phenomena not before seen in low dimensions. These dynamics are discussed in Sects.~\ref{sec:MI}-\ref{sec:quantumMI}.

\section{Three-dimensional attractive condensates} \label{sec:3d}
Attractive gases in three dimensions are prone to collapse. However, they can be stabilized in conventional harmonic traps~\cite{ruprecht1995time}. Trap potentials can be created by magnetic or far-off-resonant optical fields and are typically well approximated by a $z$-axisymmetric harmonic form $ V(r,z)=m(\omega_{r}^{2}r^{2} +\omega_{z}^{2}z^{2})/2$ with the radial (axial) trap frequency denoted by $\omega_{r}$ ($\omega_{z}$). This adds to the energy functional of Eq.~\eqref{eq:energyfunc} a potential energy term, which scales as $L^2$. A stable solution can be found within a finite parameter range, corresponding to $N\left| a_{s} \right|/\bar{l} < k_{c}$. Here, $\bar{l}=\sqrt{\hbar/(m \bar{\omega})}$ is the harmonic oscillator length, $\bar{\omega}=(\omega_{r}^{2}\omega_{z})^{1/3}$ is the geometric mean of the trap frequencies, and $k_{c}$ is a dimensionless critical value that has been computed theoretically~\cite{ruprecht1995time, houbiers1996stability,eleftheriou2000instability,gammal2001critical}.

An attractive condensate with an atom number $N>N_{c}=k_{c}\bar{l}/|a_{s}|$ tends to collapse to increase its peak density, leading to elevated three-body recombination losses. Three-body recombination is an inelastic collision process in which two atoms combine to form a dimer and a third atom helps carrying away the released binding energy to satisfy the energy-momentum conservation law. Both the atom and the dimer gain sufficient kinetic energy to leave the trap, causing atom loss.
A three-body recombination loss rate scales cubically with respect to the density, presenting itself as the dominant dissipation mechanism during a condensate collapse.

\subsection{Stability of trapped gases}\label{sec:stability}
The first observation of attractive BECs was reported by the Rice group using $^{7}\rm{Li}$ atoms~\cite{bradley1995evidence,bradley1997bose}. The combination of quadrupole and dipole magnetic fields, generated by permanent magnets, created a nearly symmetric harmonic trap with an asymmetry parameter of $\lambda = \omega_{z}/\omega_{r}\approx0.9$. Atoms were evaporatively cooled below a critical temperature for BEC at a fixed negative scattering length. Through in-situ density imaging, the existence of a limited number of condensate atoms was confirmed.

The JILA group studied the stability condition using a nearly pure BEC of $^{85}\rm{Rb}$ atoms~\cite{roberts2001controlled}. A repulsive BEC was first prepared in a cigar-shaped potential ($\lambda \approx0.4$) generated by an Ioffe-Pritchard magnetic trap. Using a magnetic Feshbach resonance, the scattering length was slowly ramped to different attractive values to induce collapse. From the remaining atom number, the dimensionless critical value was determined to be $k_{c} \approx 0.46\pm0.06$, lower than the predicted $k_{c} = 0.57$ for a spherically symmetric trap \cite{ruprecht1995time, houbiers1996stability,eleftheriou2000instability} and the expected $k_{c}=0.55$ for $\lambda=0.4$~\cite{gammal2001critical}. The contributions of finite temperature and dynamical effects were deemed insignificant~\cite{roberts2001controlled,gammal2001critical}. It was suggested that the deviation was due to high-order nonlinear effects and beyond mean-field effects~\cite{gammal2001critical,savage2003bose}.

More recently, the Chicago group demonstrated collapse experiments with 49 BECs in parallel~\cite{huang2024two}. For the preparation, a single BEC of $^{133}$Cs atoms was adiabatically loaded into a 2D array of pancake-shaped traps with $\lambda \approx 27$. Those traps were created by a combination of an oblate optical trap for confinement in the $z$ direction and $7\times7$ optical wells in the $x$-$y$ plane patterned by a DMD using similar techniques discussed in Sect.~\ref{sec:box}. The interaction of BECs was then suddenly quenched from repulsion to attraction by a Feshbach resonance. By probing the atom number in each trap and at different scattering lengths, the critical value was found to be $k_{c}=0.51\pm0.07$, which was close to the JILA group's result. The value was nevertheless significantly higher than what has been predicted for a pancake-shaped geometry ($\lambda \gg 1$)~\cite{gammal2001critical}. 

Furthermore, attractive gases can also be made transiently more stable by introducing kinetic energy during the release from a trap. The JQI group studied the stability of an expanding $^{87}\rm{Rb}$ BEC after quenching to attractive interactions using a narrow Feshbach resonance~\cite{compton2012dynamically}. A dynamically slowed collapse was observed.

\subsection{Collapse dynamics}\label{sec:collapse}
When a condensate collapses, atoms flow into the peak density region (the `singularity') until significant loss occurs. A collapse with finite atom number (energy) loss through the singularity is categorized as \emph{strong collapse} while \emph{weak collapse} refers to vanishing atom (energy) loss into the singularity~\cite{zakharov1986quasiclassical}. Both dynamics have been studied in 3D BEC experiments.

\subsubsection{Strong collapse}
Growth of an attractive BEC and its collapse dynamics near the critical condition were investigated early on by the Rice group~\cite{sackett1999measurements,gerton2000direct}. The experiment started with overcooled thermal atoms with little initial condensate population, which was removed by an energy-selective two-photon transition~\cite{gerton2000direct}. Subsequently, spontaneous condensate population growth occurred due to collisions between thermal atoms. The growing BEC intermittently collapsed near the critical condition, leading again to population depletion. The growth and subsequent collapse were observed for a few cycles, while revealing the stochastic nature of BEC collapse near the critical number.

Condensate collapse dynamics with atom number far greater than the critical value was investigated by the JILA group~\cite{donley2001dynamics}, revealing an intricate collapsing and exploding dynamics dubbed as \emph{Bosenova}. Following initial BEC production in a cigar trap, the atomic interaction was quenched from repulsion to unstable attraction and held for a variable time. Because the BEC size was smaller than the optical resolution ($\sim 7~\mu\mathrm{m}$), the group inferred the property of the collapsed gas by quenching the interaction back to a positive value for a short period of mean-field expansion before imaging the atoms. The collapse dynamics started with a brief period of no apparent atom loss nor size change before the condensate started to implode. Trap focusing tricks were implemented to search for atoms expelled from the condensate. Besides missing atoms from three-body recombination, anisotropic bursts of atoms from the collapsing condensate were observed. Elongated, seemingly low-energy `jets' along the radial direction, unlike the bursts, were also observed during the period of atom loss. The jets were speculated to originate from highly anisotropic density spikes (narrower radially) developed in the imploding condensate. Remnant atoms were observed to survive in the trap long after the collapse, sometimes with atom number greater than $N_{c}$. A later study~\cite{cornish2006formation} revealed that large remnants were formed by multiple solitary waves. 

The discovery of Bosenova triggered a series of theoretical studies employing various approaches, one of which was based on the GPE with a phenomenological three-body dissipation term with quintic nonlinearity~\cite{savage2003bose,saito2001intermittent, saito2001power, saito2002mean, santos2002collapse, adhikari2002mean}. The ANU group performed a detailed characterization of the three-body loss coefficient, and showed that the atom loss dynamics in their $^{85}{\rm{Rb}}$ condensate, measured in a similar Bosenova experiment, was well reproduced by the GPE~\cite{savage2003bose}. The underlying mechanism of Bosenova was interpreted through numerical simulations, for example, in Refs.~\cite{saito2001intermittent, saito2001power, saito2002mean}. The origin of the atomic bursts was attributed to the release of kinetic energy in local density spikes following implosion. The directional jets, on the other hand, were seen as a result of interference between coherent matter waves emerging from two adjacent spikes. 

\subsubsection{Weak collapse}
Collapse in 3D BECs introduced in the previous section was classified as strong collapse, in which a finite fraction of atoms was concentrated into a density spike. The resulting atom losses increased with attractive interaction strengths, as observed in Refs. \cite{donley2001dynamics,cornish2006formation}. In contrast, weak collapse, first discussed by Zakharov and Kuznetsov in Ref.~\cite{zakharov1986quasiclassical}, was never observed in 3D harmonically trapped gases. It was predicted that a collapsing BEC, in the absence of a trapping potential, can take an asymptotic self-similar form $|\psi|^2\rightarrow 1/r^{2}$. As the peak density diverges, the number of atoms collapsing into the singularity tends to be smaller as attraction increases. This leads to an opposite trend when comparing with strong collapse, where a stronger attractive interaction results in less atom loss in a single collapse event. The dynamics of weak collapse and the scaling of fractional atom loss were studied in the 3D GPE with a three-body loss term in Refs.~\cite{berge2002collapsing, morris2025scaling}.

The Cambridge group utilized an optical box potential to study weak collapse in uniform 3D condensates \cite{eigen2016observation}. A repulsive $^{39}{\rm{K}}$ BEC was prepared in a cylindrical box, and the scattering length was ramped to various negative values. The remaining atoms were detected via time-of-flight (TOF) absorption imaging. When the post-quench scattering length was just below a critical value, a sudden jump of atom loss was observed, indicating a single collapse event. The loss decreased with increasing $|a_{s}|$, which is consistent with weak collapse. When the scattering length became more negative than the critical value, the atom number appeared to gradually decrease until reaching a stable value. This was attributed to the occurrence of multiple collapse events until the gas regained stability. Using BECs in an optical box of tunable sizes, the Cambridge group experimentally characterized the scaling behaviors of weak collapse dynamics. Recent numerical simulations showed good agreement with the experiment~\cite{morris2025scaling}, and established an extended understanding of weak collapse.

\section{Quasi-one-dimensional attractive condensates}\label{sec:1d}
Optical or magnetic fields can be utilized to form highly elongated $z$-axisymmetric potentials ($\lambda<1$) with the tight harmonic oscillator length $l_{r}=\sqrt{\hbar/(m\omega_{r})}$ smaller than the healing length $\xi$. Such traps are often referred to as optical or magnetic waveguides, which reduce the dimensionality of trapped gases from 3D to nearly 1D. In the 1D limit, the strong radial confinement keeps atomic motion in the harmonic ground state. Using a Gaussian ansatz with the width $l_{r}$ for the radial wavefunction, we can derive the 1D GPE for the evolution of the axial mean-field wavefunction $\varphi (z,\, t)$ as
\begin{eqnarray}
i\hbar \frac{\partial \varphi}{\partial t} =  \left[ - \frac{\hbar^{2}}{2m} \frac{\partial^{2} }{\partial z^{2}} + g_{\rm{1D}} \left| \varphi \right|^{2} + V(z)  \right] \varphi \,,  \label{eq:1dGPE}
\end{eqnarray}
where $g_{\rm{1D}}= g/(2\pi l_{r}^{2})=2\hbar \omega_{r} a_{s}$ is the 1D interaction parameter and $V=m\omega_{z}^{2}z^{2}/2=m\lambda^{2}\omega_{r}^{2}z^{2}/2$ is the axial harmonic trap. When deriving Eq.~\eqref{eq:1dGPE}, we note an important assumption $l_r \gg |a_s|$ so that the two-body collisions remain 3D in nature and are free from confinement-induced effects~\cite{olshanii1998atomic}. 

One of the primary studies on attractive gases in a waveguide is the production of bright solitons, which are self-bound states that, strictly speaking, exist only in the 1D limit in the absence of any external potential. The wavefunction of a soliton containing $N$ atoms is analytically represented by a hyperbolic function $\varphi(z,t)=(\sqrt{N/2\sigma_z}) {\rm{sech}}(z/\sigma_z)\exp(-i\mu t/\hbar)$ with the width, $\sigma_z=2\xi$, set by the healing length $\xi=\hbar^{2}/(mN|g_\mathrm{1D}|)$ and $\mu=-\hbar^{2}/(8m\xi^{2})$ is the chemical potential.

However, realistic experiments are performed in a quasi-1D regime, characterized by a finite value of $l_{r}/\xi=2N |a_{s}|/l_{r}$. The underlying 3D character makes a quasi-1D condensate unstable when\footnote{For 3D BECs, $k_{c}$ is often defined in conjunction with the geometric mean of the harmonic oscillator length $\bar{l}=\lambda^{-1/6} l_{r}$, as shown in Sect.~\ref{sec:3d}.}
\begin{equation}
N>N_{c} = \lambda^{-1/6}k_{c}\frac{l_{r}}{|a_{s}|}  \,, \label{eq:3D}
\end{equation}
where $\lambda^{-1/6}k_{c} \approx 0.68$ for cigar-shaped traps with $\lambda \lesssim 0.1$~\cite{golde2018metastability,gammal2001critical}. Nevertheless, a stable solution can be found below this collapse threshold, even in a quasi-1D system with an axial harmonic trap. This presents `solitary waves,' displaying not only characters of 1D solitons but also rich dynamics arising from the inherent 3D nature of the condensate wavefunction. Single-component solitary waves have been experimentally studied by various groups: LKB in Paris~\cite{khaykovich2002formation}, Rice~\cite{strecker2002formation,nguyen2014collisions,nguyen2017formation}, JILA~\cite{cornish2006formation}, Stanford~\cite{medley2014evaporative}, ANU~\cite{mcdonald2014bright,everitt2017observation}, Paris-Sud~\cite{lepoutre2016production}, Durham~\cite{marchant2013controlled,marchant2016quantum,wales2020splitting}, Tokyo~\cite{chen2019dynamical}, Ljubljana~\cite{mevznarvsivc2019cesium}, and Strathclyde~\cite{di2019excitation}. We note that the terms `solitons' and `solitary waves' are used interchangeably in the quantum gas literature and in the following discussions.

\subsection{Formation of a solitary wave}
The formation of a single bright soliton in a quasi-1D setting was first demonstrated by the LKB group using $^{7}\rm{Li}$ atoms~\cite{khaykovich2002formation}. A localized BEC was initially prepared under a repulsive interaction. It was then released under an attractive interaction in an optical waveguide superimposed with an expulsive axial magnetic potential. The condensate was observed to propagate over a macroscopic distance of 1.1 millimeters without dispersion. Formation of the first $^{39}$K bright soliton in an optical waveguide, close to the collapse threshold, was produced in a similar fashion by the Paris-Sud group~\cite{lepoutre2016production}. An alternative method for creating a solitary wave was demonstrated by the Stanford group~\cite{medley2014evaporative}. In this experiment, evaporative cooling of $^{7}{\rm{Li}}$ atoms was performed directly under an attractive interaction. The initial atomic positions were pinned by two focused optical dipole traps intersecting with a magnetic waveguide. This method demonstrated deterministic generation of two solitary waves upon releasing condensed atoms to the quasi-1D waveguide. 

Controlled formation of solitary waves found various important applications. The ANU group demonstrated a solitary-wave Mach-Zehnder interferometer~\cite{mcdonald2014bright} based on two-photon Bragg transitions for splitting and recombining solitons in two opposite momentum states. The group found that non-dispersing solitary waves gave a sharp optimal fringe visibility relative to the dispersing wavepackets propagating under zero or repulsive interactions. Using tightly focused optical dipole potentials, the Durham group studied reflection of a solitary wave off a repulsive barrier~\cite{marchant2013controlled} and also its quantum reflection from a narrow attractive potential~\cite{marchant2016quantum}. The group further demonstrated a solitary wave `beam splitter' based on a narrow repulsive barrier approaching the width of a soliton~\cite{wales2020splitting} (see also Refs.~\cite{billam2012bright,cuevas2013interactions} and references therein). These studies opened up potential applications, for example, in bright-soliton atom interferometers.

\begin{figure}[t!]
\centering
\includegraphics[width=1.0\textwidth]{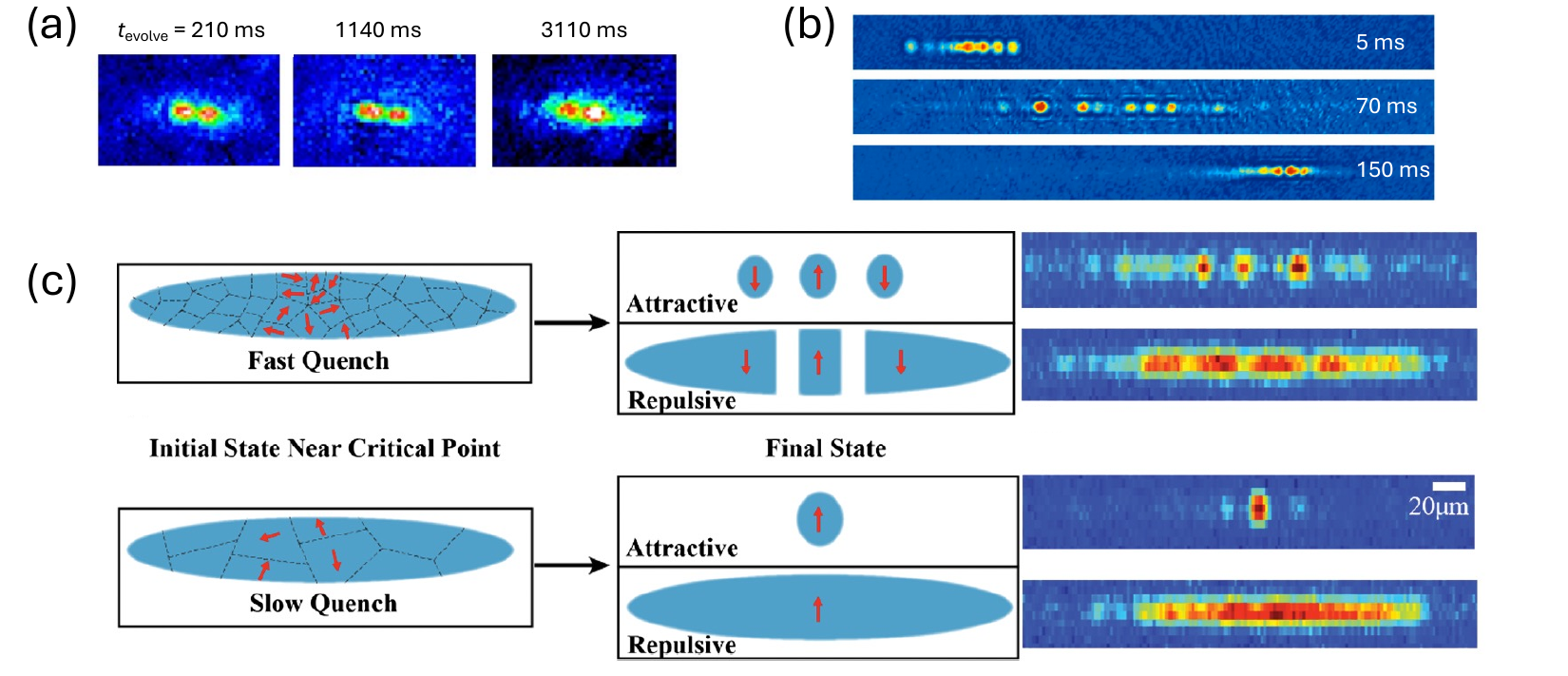}
\caption{Formation of a solitary wave train via interaction tuning in (a) the JILA experiment, adapted from Ref.~\cite{cornish2006formation}, and (b) the Rice experiment, adapted from Ref.~\cite{strecker2002formation}. (c) Solitary wave train formed via the Kibble-Zurek mechanism following a temperature quench through the BEC phase transition, observed by the Tokyo group~\cite{chen2019dynamical}. \copyright 2006 and 2019 American Physical Society (APS), and 2002 Springer Nature with permission.
}
\label{fig:fig1d_solitontrain}
\end{figure}

\subsection{Solitary wave train}
\subsubsection{Formation from an interaction quench}
A soliton train consists of an array of closely spaced solitons in a cigar trap ($\lambda<1$) as shown in Fig.~\ref{fig:fig1d_solitontrain}. It was typically observed in elongated condensates after the interaction was quickly switched to attraction, as demonstrated in harmonically trapped gases of $^{7}{\rm{Li}}$~\cite{strecker2002formation,nguyen2017formation}, $^{85}\rm{Rb}$~\cite{cornish2006formation,everitt2017observation}, and $^{133}\rm{Cs}$~\cite{mevznarvsivc2019cesium} atoms. The key formation mechanism of soliton trains, \emph{modulational instability}, will be discussed in Sect.~\ref{sec:MI}. Inside a soliton train, the stability condition Eq.~\eqref{eq:3D} indicates that $\lambda,\, l_{r}$, and $Na_{s}$ can complement each other to form solitary waves under different sets of parameters. For example, the trap geometry in the Rice experiment was $\lambda \approx 0.005$ and $\omega_{r}/(2\pi) \approx 800\,\rm{Hz}$~\cite{strecker2002formation}, while that of the JILA group was closer to a 3D shape with $\lambda\approx 0.4$ and $\omega_{r}/(2\pi) \approx 18\,\rm{Hz}$~\cite{cornish2006formation}. Since individual solitary waves can be interpreted as distinct condensates, a solitary-wave train is more stable against collapse when the atom number in each soliton does not exceed a critical value $N_{c}$. 

The situation becomes more intriguing when the solitons collide with each other, as the stability condition may be violated locally in overlapping solitons. 
However, moving soliton trains observed in experiments were surprisingly stable against collapse. 
Figure~\ref{fig:fig1d_solitontrain}(a) shows oscillations in the axial width of two solitons generated at the trap center~\cite{cornish2006formation}. Similar results were observed in~\cite{strecker2002formation}, where the spacing in a train of solitons, initially displaced from the center of an axial harmonic trap, appeared to increase when the train arrived at the trap center [Fig.~\ref{fig:fig1d_solitontrain}(b)]. These observations suggested that there was repulsive interaction between the solitons. Following Ref.~\cite{gordon1983interaction}, the apparent repulsive interaction was suggested~\cite{carr2001stability}  to originate from wavefunctions of relative phase $\pi/2 \leq \Delta \phi \leq 3\pi/2$. On the other hand, the interaction is effectively attractive for $-\pi/2<\Delta \phi < \pi/2$, allowing solitons to spatially overlap. 

\subsubsection{Formation from a temperature quench}
A soliton train can also form following a temperature quench through the Kibble-Zurek mechanism (KZM)~\cite{kibble1976topology, zurek1985cosmological}. When a system approaches the critical temperature of the BEC phase transition, the time required for a disordered normal gas to acquire phase coherence diverges in the thermodynamic limit. Therefore, when crossing the critical point by lowering the temperature in a finite time scale, the growing size of phase coherent domains will eventually `freeze' according to the temperature quench rate. Entering the BEC phase, defects will then emerge at the domain boundaries. In 1D systems with repulsive interactions, dark or gray solitons featuring phase slip across a density defect should appear spontaneously. Under attractive interactions, on the other hand, bright solitary waves are expected to form within each domain and may repel each other as a result of phase jumps across the domain boundaries. The system will form a train of solitons spontaneously, and the number of solitons should exhibit a universal scaling behavior (with respect to the temperature ramp rate) controlled by the dimensionality, the correlation length exponents and the dynamical exponent of the phase transition. 

The Tokyo group studied the BEC phase transition and the defect number scaling, under both attractive and repulsive interactions, by sympathetically cooling $^{7}{\rm{Li}}$ atoms with a $^{6}{\rm{Li}}$ thermal cloud as coolant in a quasi-1D geometry~\cite{chen2019dynamical}. In Fig.~\ref{fig:fig1d_solitontrain}(c), trains of $^{7}{\rm{Li}}$ bright (gray) solitons were observed when crossing the BEC transition at an attractive (a repulsive) interaction. The resultant number of solitons appeared to scale with the temperature quench rate. The experiment reported power-law scaling exponents which were very close in these two cases and the exponents were consistent with the predicted scaling from the KZM. This experiment confirmed that the defect scaling is insensitive to the sign of interaction, as expected from the universal scaling argument. 

\begin{figure}[t!]
\centering
\includegraphics[width=1.0\textwidth]{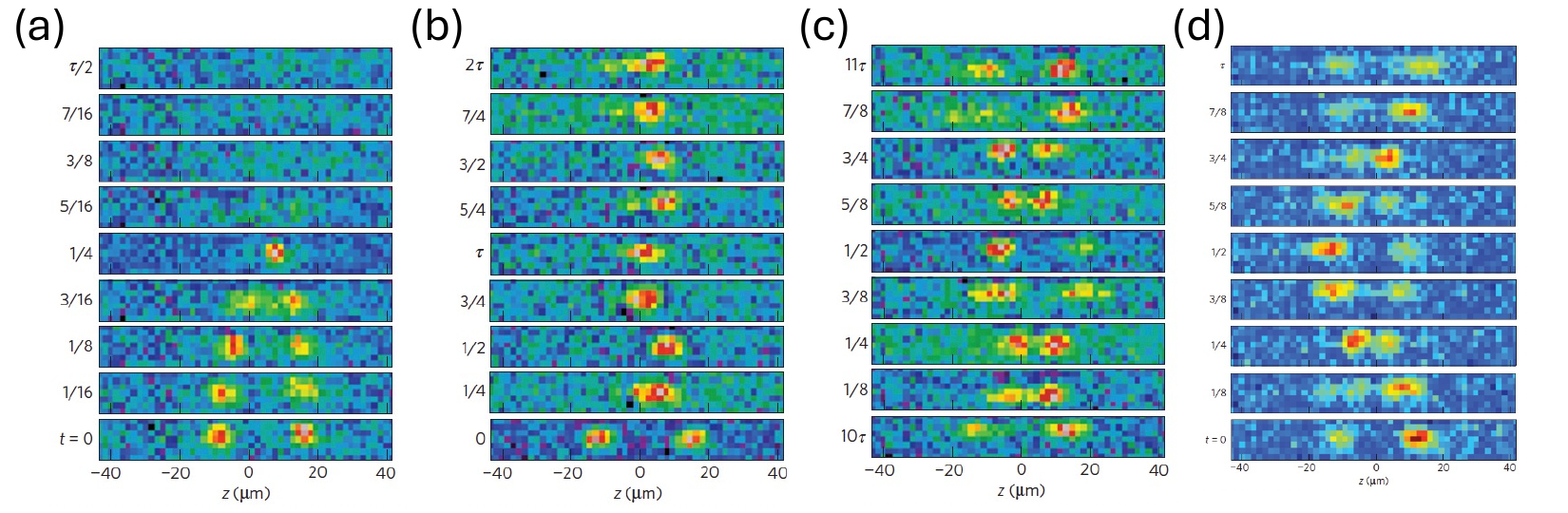}
\caption{
Collisions between solitary waves. (a), (b) In-phase collisions ($\Delta \phi=0$) resulting in soliton (b) merger or (a) supercritical collapse. (c), (d) out-of-phase ($\Delta \phi = \pi$) collisions with (c) equal and (d) imbalanced atom numbers. Adapted from Ref.~\cite{nguyen2014collisions}, \copyright 2014 Springer Nature, with permission.}
\label{fig:fig1d_solitoncollision}
\end{figure}

\subsection{Collisions between solitary waves}
In the 1D limit without an external trap, bright solitons pass through one another without changing their shape and moving speed owing to an infinite number of conservation laws~\cite{shabat1972exact}. When in close proximity, nevertheless, a pair of solitons exhibits effective interactions that depend exponentially on their separation and cosinusoidally on the relative phase~\cite{gordon1986opt}, resulting in different collision trajectories~\cite{carr2001stability,parker2008collisions}. To explore collisions of solitary waves in a quasi-1D geometry, the Rice group prepared two solitons around the center of an axial harmonic trap, which were separated by a repulsive optical barrier~\cite{nguyen2014collisions}. Once the barrier was suddenly removed, the solitons moved toward the trap center and collided with one another. The soliton motion within a single experimental run was continuously filmed via a minimally destructive polarization phase-contrast imaging. Although the relative phase between two solitary waves was random in each run, it could be inferred from the comparison between the observed soliton trajectory in a single run and the GPE simulations starting with various relative phases $\Delta \phi$. A variety of phase-dependent collision trajectories was then experimentally categorized.

For approximately in-phase collisions $\Delta \phi \approx 0$ with $N=0.53 N_{c}$ atoms in each of the solitons, a clear antinode was observed at the center amidst the collision. This typically resulted in partial collapse, reducing the atom number and oscillation amplitude of the soliton center of mass. Soliton annihilation [Fig.~\ref{fig:fig1d_solitoncollision}(a)] or merger [Fig.~\ref{fig:fig1d_solitoncollision}(b)] were sometimes observed. Out-of-phase collisions ($\Delta \phi\approx \pi$), on the other hand, resulted in the formation of a central node with zero density, displaying an effective repulsive interaction. As shown in Fig.~\ref{fig:fig1d_solitoncollision}(c), these waves were experimentally found to be robust against collapse, showing more than 20 oscillations. Importantly, the effective interactions appeared to result from destructive wave interference. The trajectory of two colliding solitons with unequal atom numbers, as shown in Fig.~\ref{fig:fig1d_solitoncollision}(d), indicated that they passed through, instead of reflected off, one another. 

\subsection{Excitation of solitary waves}
The 1D GPE supports not only the ground-state solution, i.e., the bright solitons, but also the excited states, involving fundamental breathing modes~\cite{perez1997dynamics,carr2002dynamics} and higher-order solitons~\cite{carr2002dynamics,satsuma1974b}.

\begin{figure}[t!]
\centering
\includegraphics[width=1.0\textwidth]{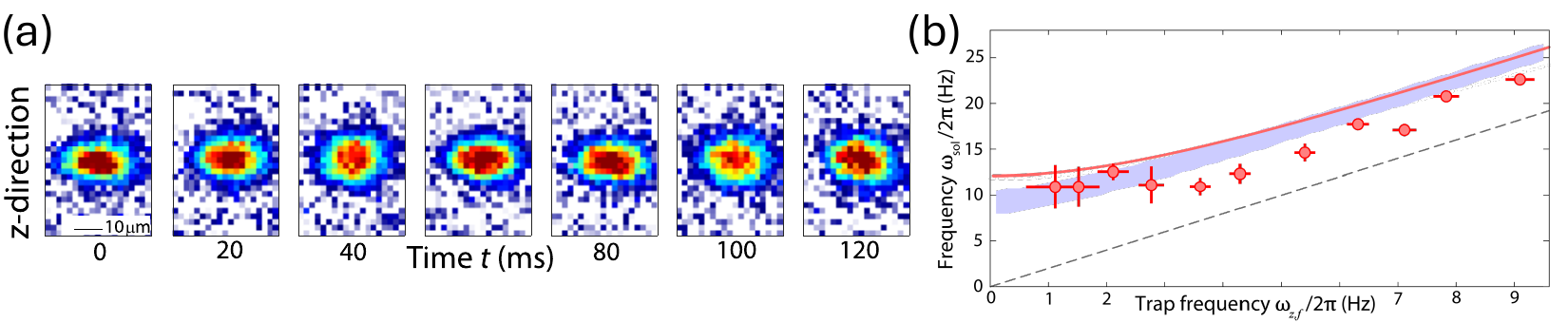}
\caption{
Breathing of a bright soliton. (a) Time-of-flight (TOF) absorption images after different hold times. (b) Breathing frequency $\omega_{\rm{sol}}$ versus the axial trap frequency $\omega_{z}$. Adapted from Ref.~\cite{di2019excitation} under the CC BY 4.0 license.}
\label{fig:fig1d_exc}
\end{figure}

\subsubsection{Fundamental breathing mode}
To see why solitons can breathe, we note that the 1D energy functional for an ansatz of the solitary wavefunction  $\varphi(z)=(\sqrt{N/2\sigma_{z}}) {\rm{sech}}(z/\sigma_{z})$ of a fixed atom number $N$ exhibits a stable local energy minimum at $\sigma_{z} = 2\xi$. The wavefunction disperses for shorter $\sigma_{z}<2\xi$ due to the excessive kinetic energy. At larger $\sigma_{z}>2\xi$, the attractive nonlinearity takes over to shrink the wavefunction, forming a self-trapping potential. An initial wavefunction that deviates from the stationary ansatz may undergo nonequilibrium oscillations around the equilibrium width $2\xi$.

To generate excited solitary waves, the Strathclyde group~\cite{di2019excitation} developed quench protocols to control the size of a quasi-1D condensate with respect to the equilibrium size of a soliton at the final attractive interaction. In these protocols, the interaction quench was combined with a rapid reduction of the axial trap frequency $\omega_{z}$ to excite the soliton breathing motions. When the initial Thomas–Fermi radius was close to the width of the expected soliton, a nearly ground-state soliton was produced, showing no significant dispersion. On the other hand, a mismatch between the two sizes led to oscillations in the soliton width, as seen in the TOF images in Fig.~\ref{fig:fig1d_exc}(a). The breathing frequency was significantly higher than that of a non-interacting gas ($2\omega_{z}$) and a repulsively interacting gas ($\sim\sqrt{3}\omega_z$). The oscillation frequency of the self-trapping potential was evaluated by extending the measured breathing frequencies to the `free soliton' limit at $\omega_{z}=0$; see Fig.~\ref{fig:fig1d_exc}(b). Comparison with an analytical calculation (red solid line) and the 1D GPE simulation (blue shaded band) showed a consistent trend.

\subsubsection{Higher-order soliton breathers}
Higher-order solitons present another intriguing type of excitations, originally predicted in the 1D nonlinear Schr\"{o}dinger equations (NLSE)~\cite{satsuma1974b}. An $M$-th order soliton, often called an $M$-soliton breather, can be interpreted as a bound state of $M$ strongly overlapping solitons with zero relative velocity and relative norm $2j-1$ for $j=1, 2, ...,M$~\cite{golde2018metastability,satsuma1974b}. The density profile of a higher-order soliton exhibits dynamical multi-peak patterns with temporal oscillations, which can be interpreted as the interference between the individual soliton components. As pointed out in Ref.~\cite{carr2002dynamics}, a matter-wave $M$-soliton breather can be generated from a fundamental bright soliton by a sudden increase of the attractive interaction strength by a factor of $M^{2}$. For example, an interaction quench by a factor of four produces an $M=2$ soliton breather composed of two constituent solitons with a norm ratio of 1 to 3. The density profile will alternate between the initial broader soliton profile and a narrow peak with two side nodes and $\pi$ phase jumps~\cite{carr2002dynamics}, arising from the interference. 

\begin{figure}[t!]
\sidecaption[t]
\includegraphics[width=0.64\textwidth]{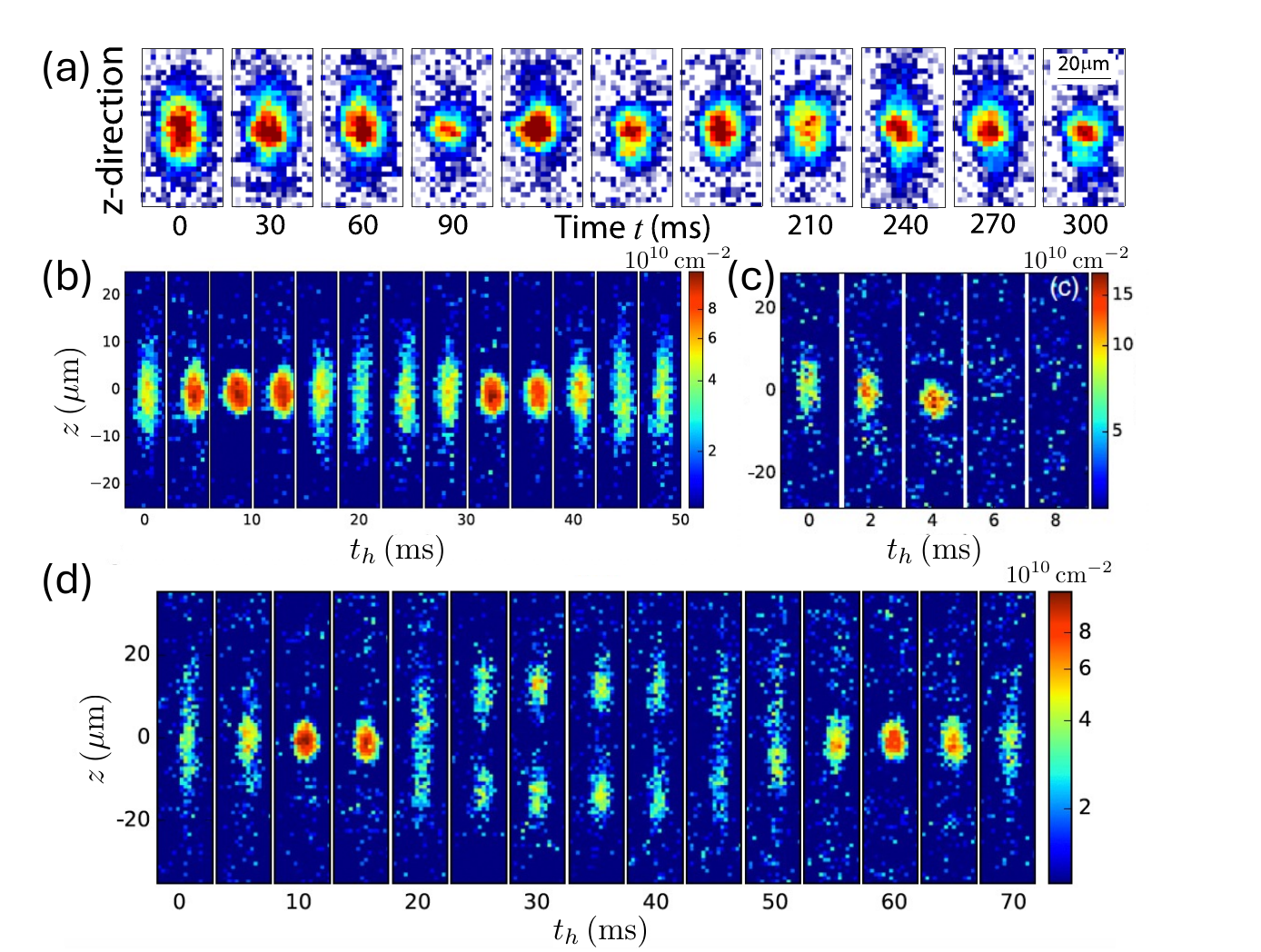}
\caption{Higher-order soliton breathers. (a) TOF images of two-soliton breathers with atom number $N< N_{c}$, adapted from Ref.~\cite{di2019excitation} under the CC BY 4.0 license. (b), (c) Non-destructive images of two-soliton breathers with (b) $N=N_{c}$ and (c) $N=1.2N_{c}$; the breather in (c) collapses at a hold time between $4$ to $6$~ms. (d) Evolution of a three-soliton breather. Panels (b)-(d) are adapted from Ref.~\cite{luo2020creation}, \copyright 2020 APS, with permission.}
\label{fig:fig1d_twosoliton}
\end{figure}

The generation of a two-soliton breather and its dynamics were demonstrated by two groups. In the Strathclyde experiment~\cite{di2019excitation}, a stable soliton was initially created at $a_s=-0.8a_0$, where $a_0$ is the Bohr radius. Following an interaction quench by a factor slightly larger than $4$ to $a_s=-4.6a_0$, a higher-order soliton was excited. The TOF density images [Fig.~\ref{fig:fig1d_twosoliton}(a)] recorded at different hold times displayed oscillations in the soliton's axial width. The measured breathing frequency was significantly lower than that of the ground-state soliton and was in very good agreement with the prediction from the two-soliton breather solution,
\begin{equation}
\omega_{B} = \frac{N^2|a_s|^2}{4l_r^2}\omega_r \approx 0.11 \left(\frac{N}{N_c}\right)^2\omega_r\,,
\label{eq:breathefreq}
\end{equation}
where the numerical factor was evaluated based on the stability condition Eq.~\eqref{eq:3D} in a cigar trap geometry $\lambda\lesssim 0.1$. In this experiment, the soliton atom number was below the estimated critical atom number of a ground-state soliton. The critical number for a two-soliton breather, denoted as $N_{c}^{(M=2)}$, is predicted to be larger than $N_{c}$~\cite{golde2018metastability}. 

The Rice group adopted an interaction quench ratio slightly below 4 to generate $M=2$ solitons. Using non-destructive imaging, the resulting breather dynamics in a single realization was continuously imaged~\cite{luo2020creation}. When $N\lesssim 1.2 N_{c}$ [Fig.~\ref{fig:fig1d_twosoliton}(b)], a two-soliton breather was observed to breathe at the expected frequency given by Eq.~\eqref{eq:breathefreq}, which was also consistent with the result of the 1D GPE simulation. For $N \approx 1.2N_{c}$, the soliton breather disappeared during the evolution, as shown in Fig.~\ref{fig:fig1d_twosoliton}(c). These suggest the collapse threshold is larger, $N_{c}^{(M=2)} > N_{c}$.

An interaction quench ratio of $(M+\alpha)^2 $ ($0<|\alpha| <  0.5$) generally produces an $M$-soliton breather. However, an excessive atom number $[1-M^2/(M+\alpha)^2]N$ will be shed out as radiation. The radiation loss would affect the breathing frequency because of the changes in the final atom number. This frequency shift was experimentally confirmed in the Rice experiment~\cite{luo2020creation}. Strong radiation shedding was also observed in the Strathclyde experiment~\cite{di2019excitation}, however, with a direct interaction quench from repulsion to attraction. Roughly two-thirds of the condensate atoms were lost after the quench, leaving behind a two-soliton breather oscillating at a frequency consistent with the value calculated using the remaining atom number in Eq.~\eqref{eq:breathefreq}.

Figure~\ref{fig:fig1d_twosoliton}(d) shows the formation of a three-soliton breather in the Rice experiment via an interaction quench ratio of $7.1$ (for $M=3$ and $\alpha\approx-0.34$). The evolution of a three-soliton breather resulted from the interference between three constituent solitons. The dynamics of the central density was found to be well fitted by the exact three-soliton breather solution~\cite{luo2020creation}.

\section{Modulational instability (MI)} \label{sec:MI}
Unlike repulsively interacting condensates that support stable phononic excitations, the wave dispersion relation $\epsilon(k)$ in attractive condensates becomes purely imaginary for long-wave excitations, that is, 
\begin{equation}
    \epsilon(k)^2 = \epsilon_k^2 + 2ng \epsilon_k < 0 \quad \text{for} \quad k<\frac{\sqrt{2}}{\xi} \,, \label{eq_disperison}
\end{equation}
where $k$ is the wavenumber of a sinusoidal perturbation, $n$ is the condensate density, $g<0$ is the interaction parameter, $\xi=\hbar/\sqrt{2m n |g|}$ is the healing length, and $\epsilon_k=\hbar^2k^2/(2m)$ is the single particle dispersion. As such, attractive condensates are modulationally unstable, hence suffering modulational instability (MI). Within the instability band, $0<k<\sqrt{2}/\xi$, any preexisting density waves can spontaneously amplify. The initial seeding can come from the zero-point (quantum) fluctuations, thermal or technical noise. In the early time dynamics of MI, the amplification rate of an unstable $k$-mode is $\sim |\epsilon(k)|/\hbar$. The modes with wavenumber $k_\mathrm{MI}\sim\xi^{-1}$ acquire the largest growth rate $\gamma=\hbar/(2m\xi^2)$. At long times, they become the dominant excitations to deplete the condensate and Eq.~\eqref{eq_disperison} is no longer valid. 

MI should occur in condensates of any dimension, with the healing length governing the length and time scales of the instability dynamics. As we shall see, the impact of dimensionality primarily manifests in the nonlinear stage of MI, when the condensate population depletes and the amplified density waves collapse. In the following sections, we discuss MI and the formation of soliton trains in quasi-1D traps and the consequences of MI dynamics in quasi-2D condensates.

\begin{figure}[b]
\sidecaption[t]
\includegraphics[width=0.64\textwidth]{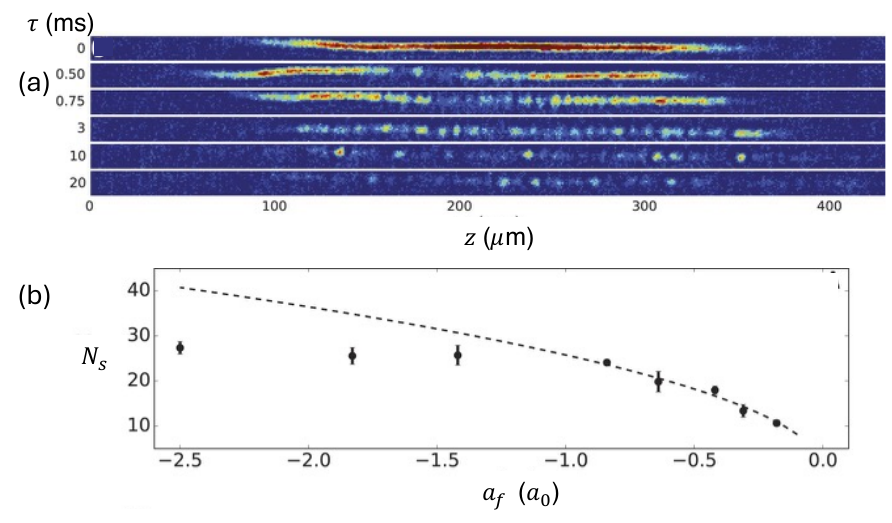}
\caption{Modulational instability in attractive quasi-1D condensates. (a) Formation of soliton trains following a hold time $\tau$. (b) Soliton number counting versus attractive scattering length. Dashed line is a simple model $N_s = \alpha R_{\rm{TF}}/(\pi\xi)$, where $\alpha =1.04$ is a fit parameter. Adapted from Ref.~\cite{nguyen2017formation}, \copyright 2017 AAAS, with permission.}
\label{fig:1dMI}     
\end{figure}

\subsection{MI and formation of soliton trains in quasi-1D traps}\label{sec:MI_1dsoliton}
It was suggested that the nonlinear dynamics of MI can cause an elongated condensate to fragment into a train of solitons, like those observed in the early 2000s~\cite{strecker2002formation}. More definite experimental proofs were later presented by the Rice group~\cite{nguyen2017formation} and the ANU group~\cite{everitt2017observation} using $^7\rm{Li}$ and $^{85}\rm{Rb}$ condensates, respectively. In these experiments, the condensates were initially confined in cigar-shaped traps with highly elongated density profiles of $\lambda\approx 0.02\sim0.1$ and $\omega_z/(2\pi)\approx 7~$Hz. The scattering length was suddenly quenched from repulsive to attractive with $-0.2~a_0 \gtrsim  a_f \gtrsim -2.5~a_0$. In the early times, no atom loss was observed as in the case of a 3D Bosenova (Sect.~\ref{sec:collapse}). After a sufficient hold time (around a few milliseconds comparable to $\gamma^{-1}$), significant density modulations appeared within the bulk (center) of the condensate [Fig.~\ref{fig:1dMI}(a)], followed by initiation of atom loss and formation of a train of solitons along the trap. The observed number of solitons in the Rice experiment was found to be in good agreement with a simple length-scale estimation: $N_s \approx 2R_{\rm{TF}}/(2\pi\xi)$, where $R_{\rm{TF}}$ is the Thomas-Fermi radius before the quench and $2\pi\xi$ is the wavelength of the most unstable mode; see Fig.~\ref{fig:1dMI}(b). The ANU experiment had similar observations. Both groups applied non-destructive phase contrast imaging techniques to track the in situ density evolution. They confirmed that the soliton trains were formed from MI seeded by initial density noise rather than the modulations from self-interference at the edge of the gases, which could be deterministically generated right after the interaction quench~\cite{carr2004spontaneous}. Through observing the breathing motion of the soliton train following the nonlinear stage of MI, the Rice experiment further confirmed that the neighboring solitons were primarily repulsively interacting. This implied that the solitons formed from MI were closer to $\pi$ out-of-phase than being in-phase with respect to each other. 

Lastly, we note that formation of a soliton train from MI was also observed in a radio frequency-coupled two-component $^{39}{\rm{K}}$ BEC~\cite{sanz2022interaction}, where a fast interaction quench was implemented by addressing the condensate atoms in one spin state with another state showing large inter-state attraction. 

\begin{figure}[b]
\sidecaption[c]
\includegraphics[width=0.64\textwidth]{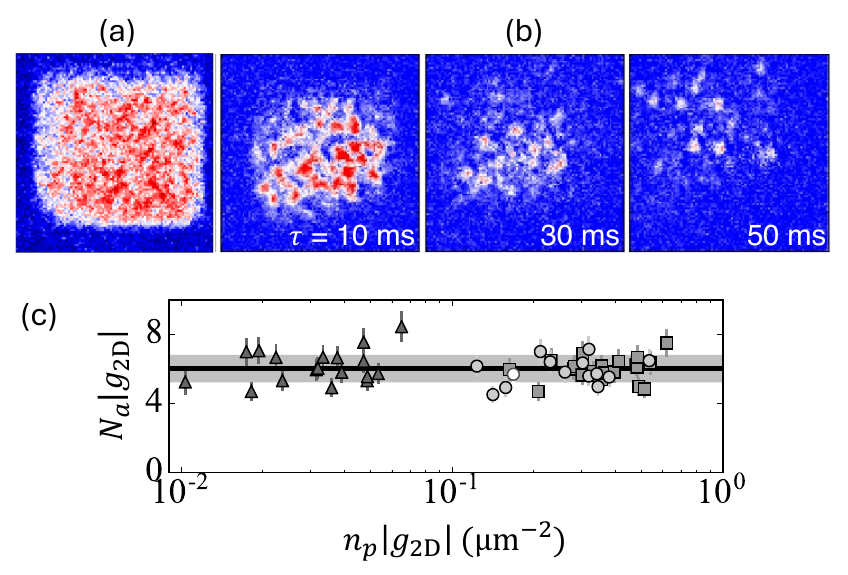}
\caption{Modulational instability in attractive quasi-2D condensates. (a) before and (b) after the interaction quench. (c) Measured $N_a|g_\mathrm{2D}|$ versus peak interaction parameter $n_p|g_\mathrm{2D}|$. Panels (a) and (b) are adapted from Ref.~\cite{Chen2020ObservationGases}; panel (c) from Ref.~\cite{Chen2021ObservationSolitons}. \copyright 2020 and 2021 APS, with permission.}
\label{fig:2dMI}      
\end{figure}

\subsection{MI and formation of Townes solitons in quasi-2D traps}\label{sec:MI_2dsoliton}
MI-induced condensate fragmentation was also demonstrated in attractive 2D condensates prepared in optical boxes. This was observed at Purdue using $^{133}\rm{Cs}$ Bose gases~\cite{Chen2020ObservationGases}. A 3D BEC was first compressed into a 2D box [Fig.~\ref{fig:2dMI}(a)] with a vertical trap frequency $\omega_z/(2\pi) \approx1.75~$kHz. 
Experiments were performed by quenching the atomic interaction from repulsion to attraction ($|a_s| \lesssim 20~a_0$, corresponding to $g_\mathrm{2D}\gtrsim -0.03$) while simultaneously removing the horizontal potential of the optical box, allowing the condensate atoms to expand freely in the 2D plane before destructive absorption imaging was performed. Following the quench, density waves grew throughout the sample, as shown in Fig.~\ref{fig:2dMI}(b), suggesting noise-seeded MI similar to the 1D case. No significant atom loss was observed before $\tau \sim \gamma^{-1}$. This early time dynamics will be discussed in detail in Sect.~\ref{sec:quantumMI}, where we carefully analyze MI-induced density noise. Entering the nonlinear stage of MI, $\tau \gtrsim \gamma^{-1}$, the density waves collapsed and the condensate fragmented into many density blobs. The observed blobs had averaged diameter $d\sim \pi \xi$, where $\xi$ was the healing length, and mean atom number $N_a \sim  6/|g_\mathrm{2D}|$ regardless of the peak density of blobs as shown in Fig.~\ref{fig:2dMI}(c). This remarkable scaling behavior could also be deduced from the length-scale argument of MI in two dimensions. The atom number in the blobs was estimated to be $\sim n d^2 \approx 5/|g_\mathrm{2D}|$, which was indeed very close to the experimental observation. Unlike in the 1D case, there are no stable 2D solitons. However, the observed mean atom number in the density blobs was very close to the so-called Townes threshold $N_\mathrm{th}= 5.85/|g_\mathrm{2D}|$, which is the norm of the only stationary solution in the 2D GPE--the Townes soliton. A Townes soliton is a self-trapped stationary solution of the 2D NLSE, first discovered in Ref.~\cite{chiao1964self}. As we shall discuss in Sect.~\ref{sec:ts}, it appeared that many of the blobs formed Townes solitons and thus appeared long-lived. 

Interestingly, following wave collapse, the experimentally measured total atom number decay displayed an unexpected two-body loss behavior~\cite{Chen2020ObservationGases} not observed in the quasi-1D case~\cite{nguyen2017formation}. This peculiar dynamics was explained by an effective two-body inelastic collision model: Following condensate fragmentation, Townes soliton-like density blobs can still collide with one another. Since there are no stable solitons in 2D, collisions trigger collapse and fast atom loss. The atom number decay thus reflects the `two-body' collision rate between blobs. The experiment measured a nearly constant soliton binary loss coefficient regardless of the interaction parameter $g_\mathrm{2D}$, which agreed very well with a simple prediction $\Gamma_s\approx \sqrt{2}\pi\hbar/m$ by applying the scaling argument of MI in two dimensions.

\subsection{Quasi-1D nonlinear dynamics of MI }
So far, experiments performed in the quasi-1D and 2D gases associated the nonlinear stage of MI with wave collapse, which led to the formation of solitary waves. Nevertheless, it is known that the nonlinear dynamics of MI in a 1D NLSE can also exhibit recurrence dynamics owing to the integrability that leads to an infinite number of conservation laws. Representative solutions include the Akhmediev breathers~\cite{akhmediev1986modulation} and the Peregrine soliton~\cite{peregrine1983water}. The former show solutions of different dynamics, evolving from a uniform wave into spatially (temporally) localized and temporally (spatially) periodic waveforms, while the latter exhibits a spatiotemporally localized peak over a uniform background. In another case, when a smooth waveform is terminated with sharp gradient(s), like a top hat, and an attractive interaction is suddenly introduced, dispersive shock waves (DSWs) will be emitted from the edge(s) of the waveform~\cite{el2016dam}. The interference of two counter-propagating DSWs from the edges will form dynamical, quasi-periodic wave modulations in the system. Each resulting peak near the center resembles a Peregrine soliton-like structure, featuring near-$\pi$ phase jumps across its core and tails (see also Chapter~5). The evolution of DSWs was previously suggested as an alternative mechanism seeding the formation of solitary wave trains in a quasi-1D gas~\cite{carr2004spontaneous}. 

The recurring dynamics of MI remains largely elusive in attractive condensate experiments. This may be partially due to random noise-seeded density modulations dominating the nonlinear dynamics of MI. In a recent study, the Purdue group explored quench dynamics of condensates in a quasi-1D box~\cite{tamura2025observation}. The samples were prepared by loading a quasi-2D gas (with $l_z \ll \xi$) into two elongated boxes, each with a transverse oscillator length $l_y\sim1.1~\mu\rm{m}$ comparable to the healing length. The condensates featured near uniform 1D density along the box and sharp density gradients at the edges, where DSWs would be emitted upon the interaction quench. Through in situ density and relative phase (obtained through matter-wave interferometry discussed in Sect.~\ref{sec:box}) correlation measurements, the samples displayed dynamical density modulations of changing periodicity as well as suppression and partial recovery of long-range phase coherence during the nonliner evolution. These behaviors were reproduced by 2D GPE simulations of quench dynamics in narrow 1D boxes. The measurements revealed the interplay between the DSWs and the noise-seeded density modulations~\cite{tamura2025observation}. We note that engineered, deterministic generation of a Peregrine soliton was recently reported in Ref.~\cite{romero2024experimental} by the WSU group (see Chapter~5), using harmonically trapped two-component quasi-1D gases of $^{87}$Rb atoms with an effective attractive interaction in the minority component (see also Sect.~\ref{sec:ts} for discussions on two-component gases). The same group also demonstrated DSWs generated after the removal of a sharp repulsive optical barrier~\cite{mossman2025nonlinear}.

\section{Quasi-two-dimensional attractive condensates}
\label{sec:2d}

\subsection{Observation of scale-invariant Townes solitons}\label{sec:ts}
We now discuss the Townes soliton, the stationary solution of the 2D NLSE~\cite{chiao1964self}, and its scale invariance. Higher order solutions with nodal points in the radial wavefunction or non-zero phase winding (vorticity) have also been identified~\cite{carr2006vortices,malomed2016multidimensional}. We consider here a wave function of the form $\psi = \phi(r)e^{iS\theta}$, where $\phi$ is localized in space and $S \in \mathbb{Z}$ is the phase winding number. Due to the scaling symmetry in the 2D GPE, all physical observables can be rescaled into dimensionless units and we arrive at a scale-invariant GPE,
\begin{equation}
- \frac{1}{2 }\left(\frac{d^2\phi}{dR^2} + \frac{1}{R}\frac{d\phi}{d R} - \frac{S^2}{R^2} \phi\right) - |\phi|^2\phi = \tilde{\mu} \phi\, , \label{eq:sc_GPE}
\end{equation}
where $R=\kappa r$ is the rescaled radial coordinate and $\tilde{\mu}=m\mu/\hbar^2\kappa^2$ is the scaled chemical potential. Each stationary solution to this 2D GPE has a distinct value of $\tilde{\mu}$ and a unique scale-invariant profile. Given a peak density $n_p$ and the interaction parameter $g_\mathrm{2D}$, the scale factor $\kappa = \sqrt{n_p|g_\mathrm{2D}|}$ can be determined and the physical solution takes the form 
\begin{equation}
\psi (r,\theta) =\sqrt{n_p} \phi(\kappa r) e^{iS\theta}\,.    
\end{equation}
A Townes soliton is the fundamental solution with zero winding number $S=0$. The norm of a Townes soliton $N_\mathrm{th} = \int |\psi (r,\theta)|^2 d \mathbf{r} = 5.85/|g_\mathrm{2D}|$ is clearly scale-independent regardless of the peak density or the physical size. A wave packet with its norm beyond the Townes threshold $N > N_\mathrm{th}$ would collapse due to excessive nonlinear attraction. If $N < N_\mathrm{th}$, the wave packet disperses indefinitely\footnote{Nevertheless, adding an external trap with $N < N_\mathrm{th}$ can yield stable solitary waves.}. As a result, Townes solitons are inherently unstable and are difficult to form in experiments.

\begin{figure}[t]
\centering
\includegraphics[width=0.8\textwidth]{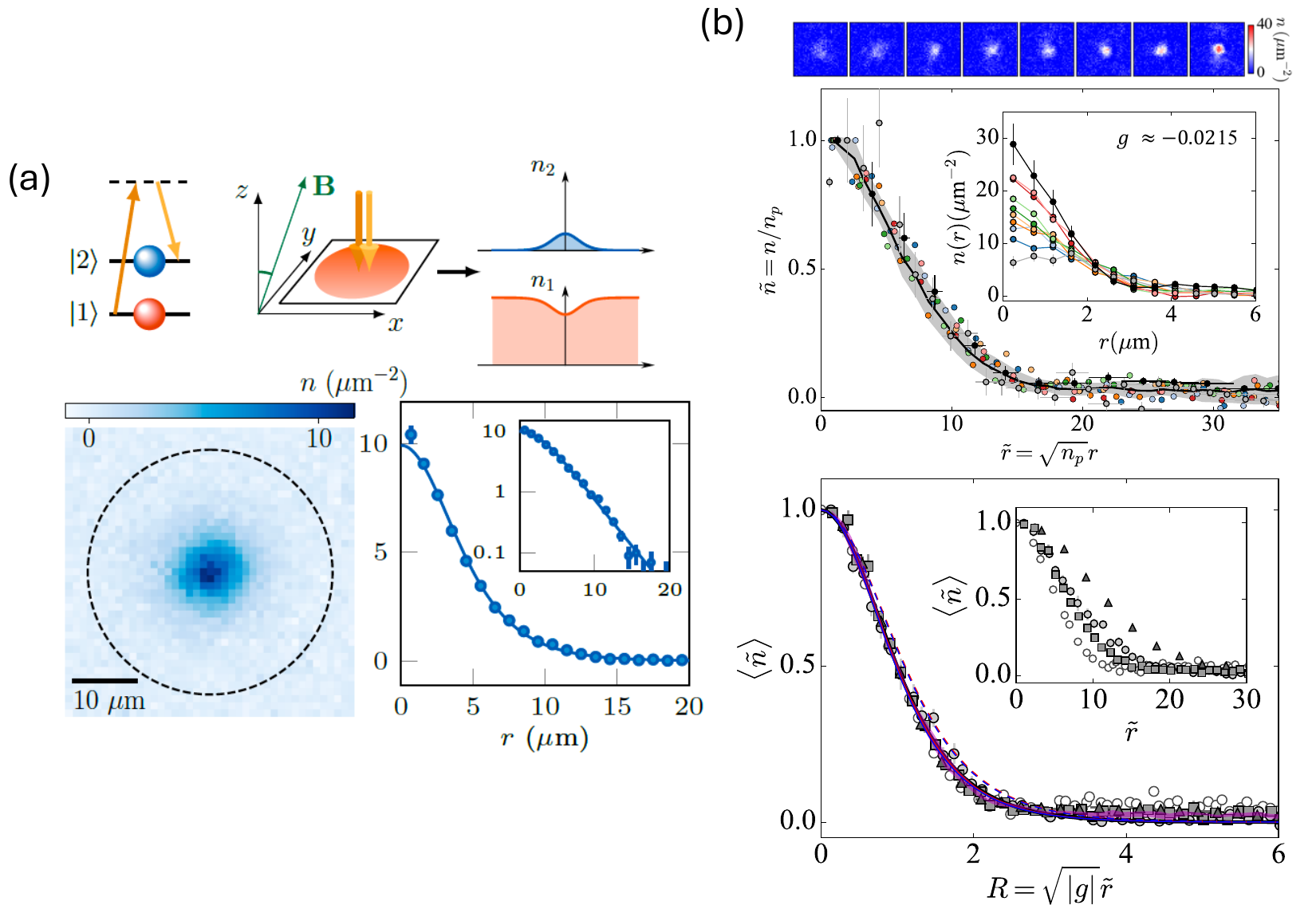}
\caption{Formation of scale-invariant Townes solitons. (a) The Paris experiment~\cite{bakkali2021realization}. Using a spatially resolved optical transfer technique, a scale-invariant Townes profile was directly imprinted in a two-component planar Bose gas. (b) The Purdue experiment~\cite{Chen2021ObservationSolitons}. Solitons were created following MI-induced wave collapse. Rescaled density profiles, measured at different peak densities and interactions, collapse onto a single curve, agreeing with the scale-invariant Townes profile. \copyright 2021 APS, with permission.}
\label{fig:TS}      
\end{figure}

Townes solitons were experimentally created at Purdue~\cite{Chen2020ObservationGases,Chen2021ObservationSolitons} and LKB in Paris~\cite{bakkali2021realization} in 2D box potentials. Unlike in one dimension, it was challenging to form 2D solitons due to the requirement to match the norm of a matter wave with the Townes threshold, which is typically smaller than a few hundred atoms, and the necessity to realize the exact stationary profile. The Paris group achieved this by directly imprinting a Townes soliton via a DMD-controlled Raman transition in a two-component $^{87}$Rb condensate; see Fig.~\ref{fig:TS}(a). In the experiment, a small fraction of atoms in the ground hyperfine state $\ket{1}=\ket{F=1,m_F=0}$ was locally transferred to another state $\ket{2}=\ket{F=2,m_F=0}$. The atoms in state $\ket{1}$ serve as a bath to mediate interaction between atoms in state $\ket{2}$. While the two-body interactions between the inter- and intra-spin components were all repulsive $g_{ij}>0$, very small differences among them resulted in an effective two-body attractive interaction $g_e = g_{22} - g_{12}^2/g_{11}<0$ for the $\ket{2}$ component, realizing an effective single-component NLSE; for more discussions, see Chapter~4. Using this method, the Paris group was able to engineer Townes soliton profiles of different sizes with fixed $N|g_e|\approx 5.85$, and to test their stationarity and scale invariance. It is also worth noting that, in the case of larger atom number beyond the Townes threshold $N_\mathrm{th}$, there is no true collapse behavior as this would lead to local depletion of bath atoms and deviation from the NLSE. The system supports localized solutions for $N>N_\mathrm{th}$, from Townes soliton-like to droplet-like states~\cite{bakkali2023cross}. 

The Purdue group, on the other hand, utilized MI to create small fragmented wave packets with atom numbers universally around $N_\mathrm{th}$ (Fig.~\ref{fig:2dMI}). The measured density profiles of different physical sizes indeed matched well with the scale-invariant Townes profile [Fig.~\ref{fig:TS}(b)]. Using various initial densities and interaction strengths, the group confirmed that scale-invariant Townes solitons on 60-fold differences in $n_p|g_\mathrm{2D}|$ can be formed from MI-induced wave collapse~\cite{Chen2021ObservationSolitons}.  

\begin{figure}[t]
\sidecaption[t]
\includegraphics[width=0.64\textwidth]{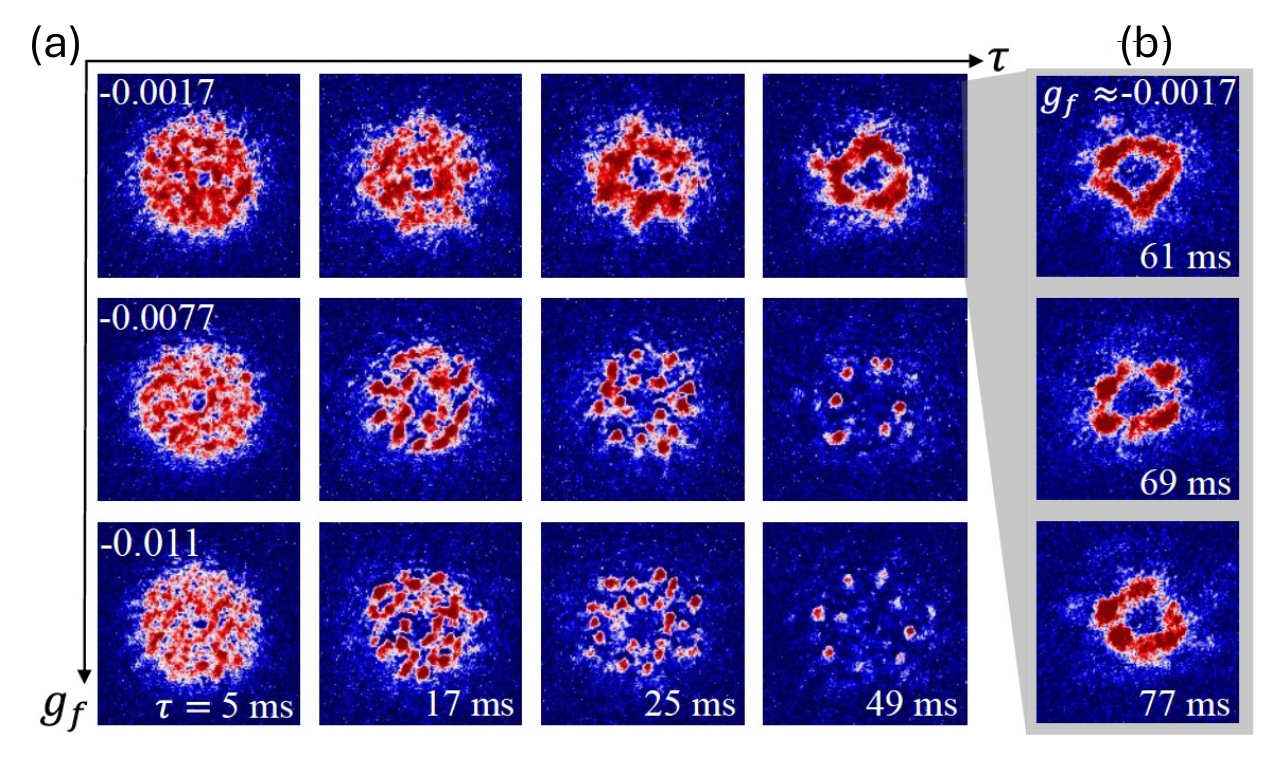}
\caption{Collapse of a quantum vortex in attractive condensates. (a) Single-shot images with variable interaction $g_f$ and 2D TOF time $\tau$. (b) Vortex soliton-like structures imaged at longer times $\tau$. Figure adapted from Ref.~\cite{banerjee2024collapse}, \copyright 2025 APS, with permission.}
\label{fig:VS}      
\end{figure}

\subsection{Vortex solitons}
A vortex soliton is a stationary solution to Eq.~\eqref{eq:sc_GPE} with non-zero winding number $S\neq 0$. Such a solution carries a phase singularity and defect at the center $r=0$, pushing the peak of the radial function to a finite radius $r\gtrsim \xi$. A fundamental vortex soliton with $|S|=1$ forms a localized, stationary ring-shaped profile. Similar to the Townes solitons, a vortex soliton is experimentally challenging to form in an attractive condensate because of the instability to collapse radially when the atom number is beyond a threshold. Moreover, the azimuthal size of a vortex soliton is longer than the most unstable length scale of MI $\sim 2\pi\xi$. As a result, even when one starts with an ideal self-trapped radial profile, a vortex soliton is still unstable against azimuthal long-wave perturbations. Following a Bogoliubov analysis~\cite{saito2002split,banerjee2024collapse} similar to that carried out for Eq.~\eqref{eq_disperison}, azimuthal modulations with a discrete mode number $l=2$ have been predicted to possess the largest imaginary frequency. A vortex soliton in free space is therefore expected to split due to the azimuthal MI. In Ref.~\cite{mihalache2006vortex}, it was found that adding a harmonic trapping potential can stabilize an $S=1$ vortex soliton.

Thanks to the application of an optical box, a vortex soliton-like structure has been experimentally observed at Purdue following the collapse of a quantum vortex at an attractive interaction~\cite{banerjee2024collapse}. The experiment followed a protocol developed in Refs.~\cite{samson2016deterministic,gertjerenken2016generating} to sweep a DMD-controlled `chopstick' potential across a circular superfluid and deterministically generated a singly-charged vortex at the center. The atomic interaction was ramped to attraction while the horizontal box wall was removed in a few milliseconds, just slow enough to avoid creating other excitations such as DSWs. The group discovered that a condensate of initial radius $r \sim 2\xi$ can radially collapse onto a vortex soliton-like, quasi-stationary density profile at time $\tau \sim \gamma^{-1}$ (Fig.~\ref{fig:VS}), similar to the wave collapse timescale found in MI. For samples with $r \gtrsim 3\xi$, more than one density ring formed radially [second and third rows of Fig.~\ref{fig:VS}(a)]. Furthermore, the group identified that azimuthal MI fragmented all samples at time $\tau\gtrsim 2\gamma^{-1}$ into disordered, but roughly circular arrays of Townes soliton-like wave packets. 

A 2D GPE wavefunction, when seeded with random Gaussian noise to model the initial quantum phonon fluctuations, qualitatively reproduced the time scales of radial and azimuthal wave collapse as well as the formation of vortex soliton and Townes soliton-like profiles~\cite{banerjee2024collapse}. However, it produced amplified density fluctuations of much greater magnitude than those observed in the experiment. This discrepancy was attributed to effects beyond the mean-field approximation. 

\section{Quantum dynamics in modulational instability}\label{sec:quantumMI}
 The quench experiments discussed in Sects.~\ref{sec:MI} and \ref{sec:2d} revealed the classical nonlinear wave nature of attractive condensates, and quantitatively associated the observed soliton formation dynamics with the length and time scales of MI. Due to the initial ultracold temperature, MI-induced density modulations can be predominantly amplified from quantum fluctuations in an attractive condensate. This suggests that, when dissipation and dephasing effects were not yet significant, the amplified density noise could carry non-classical correlation signatures. A very useful tool--the density noise power spectrum--can be devised to analyze the dynamics of density noise. This new observable can further reveal important spatiotemporal dynamics and quantum correlations resulting from MI. Therefore, we devote this section to a detailed analysis of such a tool. The following example was carried out in a quasi-2D experiment but the method can be extended to one or even three dimensions. 

\subsection{Density noise power spectrum}
To measure the density noise power spectrum, many atomic samples should be prepared under identical conditions. Destructive absorption imaging can be performed to record each sample's density distribution $n(x,y)$ in a \emph{single shot}. Several dozen density images suffice to evaluate the density noise power spectrum, 
\begin{equation}
    S(\mathbf{k}) = \frac{\langle |\delta n(\mathbf{k})|^2\rangle}{N} \,, \label{eq:power_spectrum}
\end{equation}
where $\delta n(\mathbf{k})=\delta n(k_x,k_y)$ is the 2D Fourier transform of the shot-to-shot density noise\footnote{If the samples are uniform gases, the density noise power spectrum can be approximated by the density power spectrum $S(\mathbf{k})=\mean{|n(\mathbf{k})|^2}/N$ since $\delta n(\mathbf{k}) \approx n(\mathbf{k})$ for spatial frequency $k_{x,y} \gg L^{-1}_{x,y}$, where $L_{x,y}$ is the linear size of the samples.} $\delta n(x,y) = n(x,y)-\langle n(x,y)\rangle$, the symbol $\langle\dots\rangle$ denotes averaging over an ensemble of samples, and $N$ is the mean atom number. When there is no expected anisotropy, the power spectrum can be azimuthally averaged for every $|\mathbf{k}|=k$, giving $S(k)$. A peak in $S(k)$ indicates that periodic density modulations around a spatial frequency $k$ are present in the samples, which is the primary signature of MI. 

The atomic signal is recorded within a spatial frequency range permitted by the diffraction limit, $k\leq 2\pi \rm{NA}/\lambda_{0}$, where $\lambda_{0}$ is the transition wavelength and $\rm{NA}$ is the numerical aperture of the imaging optics. Moreover, the spectrum may be distorted by aberrations in the imaging system\footnote{The measured power spectrum should also contain a white noise floor, which is contributed by the photon shot-noise and the pixel-wise readout noise from images recorded on a CCD camera. This noise floor can be subtracted off.}. This effect can be fully characterized and corrected to recover the unaberrated signal. For interested readers, the calibration procedures are detailed in Refs.~\cite{Chen2020ObservationGases,hung2011extracting}. In typical experiments, an imaging system with moderately high $\rm{NA} \sim 0.5$ can detect modulation wavenumber up to $k \sim 4/\mu\rm{m}$ or wavelengths down to $\sim 1.5~\mu\rm{m}$. The resolving power of typical imaging systems limits the studies of MI to weak nonlinearities, typically at $k^{-1}_\mathrm{MI}=\xi \gtrsim 2~\mu\rm{m}$. 

\begin{figure}[t]
\centering
\includegraphics[width=1\textwidth]{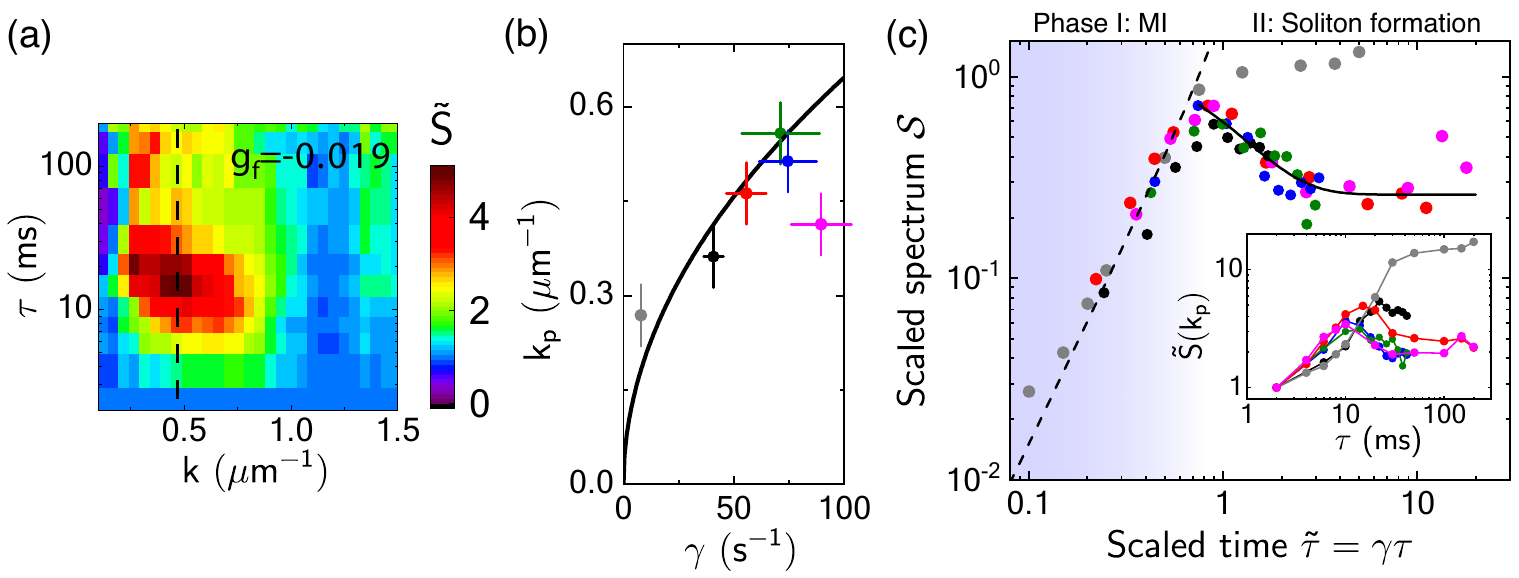}
\caption{Density power spectrum from Ref.~\cite{Chen2020ObservationGases}. (a) The spectrum peaks at the most unstable mode $k_\mathrm{p}$ (dashed line). (b) Measured $k_\mathrm{p}$ for samples of different interaction energies (color symbols) shows agreement with $k_\mathrm{MI}$ (solid curve). (c) Scale power spectra $\mathcal{S}$ (color symbols) of the corresponding samples in (b) display a universal dynamics for MI and wave collapse for soliton formation. \copyright 2020 APS, with permission.}
\label{fig:spectrum}     
\end{figure}

Figure~\ref{fig:spectrum} displays time evolution of the density power spectra measured in the Purdue experiment. A quench to an attractive interaction takes place at $\tau=0$. The spectrum in panel (a) clearly demonstrates the growth of density noise within an instability band, peaking at the expected most unstable wavenumber $k_\mathrm{p} \sim \xi^{-1}$ after a short time; see panel (b). When properly rescaled, the power spectra measured with condensates at different attractive interactions displayed a universal behavior [panel (c)]. The rescaled spectra showed a growth era that is consistent with a hyperbolic function (dashed line) predicted from a linear analysis of MI (see also the following section), 
\begin{equation}
\mathcal{S}(\tau)=\beta \left[\frac{S(k_\mathrm{p},\tau)}{S(k_\mathrm{p},0)} - 1\right] = 2 \sinh^2(\gamma\tau) \,, \label{eq:sinh}
\end{equation}
where $\beta$ is a rescaling factor controlled by the interaction parameter $g_\mathrm{2D}$ before and right after the quench~\cite{Chen2020ObservationGases}. Beyond time $\tau\gtrsim \gamma^{-1}$, the amplitude of the spectrum declined as an indication of wave collapse and soliton formation.  

\subsection{Quasiparticle pair production}\label{sec:quasi-pair}
In the following, we first discuss the microscopic quantum dynamics in MI followed by an experimental characterization of quantum correlations developed in the attractive condensates. To gain insight into how quantum noise is amplified, we focus on the early time (linearized) dynamics of MI right after the interaction quench, where the modulation amplitudes are small and the condensate is primarily occupied by atoms in the zero-momentum state. We consider the dynamics of quantized collective excitations, i.e., quasiparticles, and follow a modified Bogoliubov transformation to analyze the Hamiltonian governing the excitations~\cite{Chen2020ObservationGases,feng2018coherent}. Assuming the zero momentum state is macroscopically occupied (population $N_0\approx N\gg 1$), the effective Hamiltonian for quasiparticles in wavevectors $\pm \mathbf{k}$ satisfying $\epsilon(k)^2<0$ in Eq.~\eqref{eq_disperison} can be approximated as
\begin{equation}
    \hat{H}_\mathbf{k}\approx \epsilon (k)\left(\hat{b}_\mathbf{k}^\dagger\hat{b}_{-\mathbf{k}}^\dagger+\hat{b}_\mathbf{k}\hat{b}_{-\mathbf{k}}\right) \quad \text{for} \quad 0<|\mathbf{k}|<\frac{\sqrt{2}}{\xi} \,, \label{eq:pair}
\end{equation}
where $\hat{b}_\mathbf{k}^{(\dagger)}$ is the quasiparticle annihilation (creation) operator. Albeit valid only transiently, this effective Hamiltonian indicates that quasiparticles are created (annihilated) in pairs while removing (creating) two atoms in the ground state.

The pair-production dynamics, which is induced by the cubic nonlinear attraction, is analogous to the spontaneous four-wave mixing (SFWM) process in a nonlinear $\chi^{(3)}$ medium, which is a well-known scheme in quantum optics to create squeezed light and entangled photon pairs~\cite{slusher1985observation,du2008narrowband,cohen2009tailored,luo2022quantum}. In the SFWM, two photons in a classical pump field interact with a nonlinear medium to spontaneously emit a pair of time-frequency entangled photons, which is enforced by energy and momentum conservation. This spontaneous process is seeded by electromagnetic vacuum fluctuations. In an attractive condensate undergoing MI, on the other hand, pair production of quasiparticles can be triggered by quantum fluctuations, thermal and technical noise in the condensate wavefunction. The emitted quasiparticles remain overlapped with the condensate and can continue to stimulate new particles or annihilate in pairs as described by Eq.~\eqref{eq:pair}. This peculiar dynamics leads to the growth function of the noise power spectrum in Eq.~\eqref{eq:sinh}, which agrees well with the observation in Fig.~\ref{fig:spectrum}. 

We note that the early time MI dynamics is also akin to the dynamical instability in $F=1$ (or $2$) spinor condensates, where spin-changing collisions force two atoms in the $\ket{m_F=0}$ condensate to populate the $\ket{m_F=\pm 1}$ states in vacuum. This effect enabled the observation of parametric amplification of vacuum fluctuations~\cite{chang2004observation,leslie2009amplification,klempt2010parametric}, spin squeezing~\cite{gross2011atomic,lucke2011twin}, entanglement distribution and Einstein-Podolsky-Rosen steering of BECs~\cite{kunkel2018spatially,fadel2018spatial,lange2018entanglement}.

\subsection{Characterization of quantum entanglement}
\begin{figure}[t]
\sidecaption[t]
\includegraphics[scale=.34]{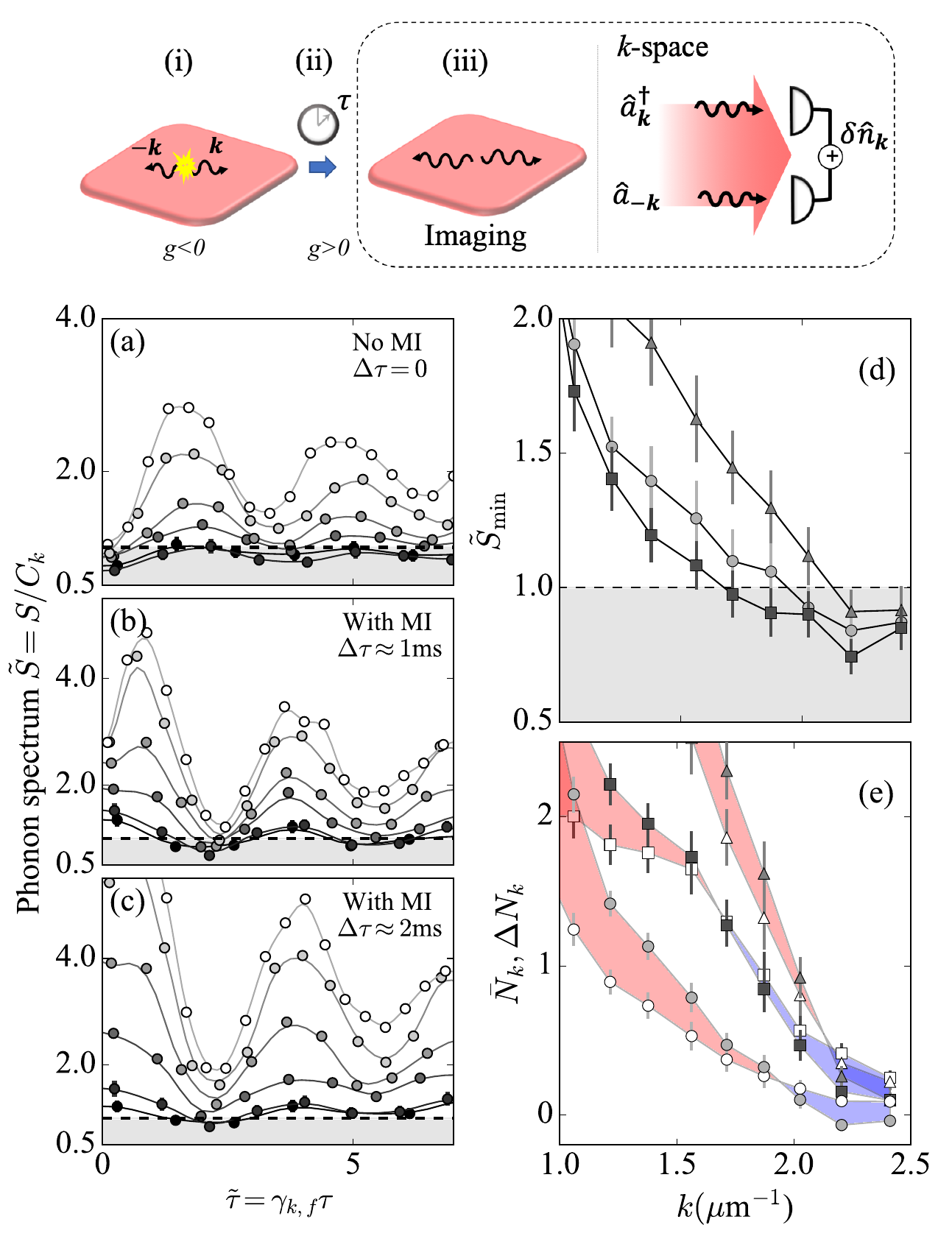}
\caption{Quasiparticle pair production and detection. Experimental scheme (top panel): Interaction quench to $g<0$ for time $\Delta\tau$ (i) is followed by a second interaction quench to $g>0$ and hold time $\tau$ (ii). In situ density noise measurement in Fourier space (iii) is a homodyne measurement of quasciparticles with the ground state atoms. Lower panels: (a)-(c) Rescaled power spectrum for $k\approx($1.3, 1.6, 1.8, 2.1, 2.2, 2.4)$/\mu\mathrm{m}$ (bright to dark circles). (d), (e), Squeezing minima (d), mean phonon number $N_{k}$ (open symbols) and pair correlation amplitude $\Delta N_{k}$ (filled symbols) (e) are plotted for $\Delta\tau=0$ (circles), 1 (squares), 2 (triangles)~ms, respectivley. Adapted from Ref.~\cite{Chen2021ObservationInteraction}, \copyright 2021 APS, with permission.}
\label{fig:entanglement}       
\end{figure}

Experimentally, how do we confirm nonclassical correlations, that is, quantum entanglement in the MI-induced density modulations? Once again, the density noise power spectrum lends itself as a powerful tool for characterizing quantum entanglement. In essence, the observed condensate density noise is a manifestation of interference between the ground state atoms and quasiparticle excitations. Using the quantum optics terms to describe atomic density imaging, the ground state atoms act like a coherent `local oscillator' interfering with the quasiparticles that are `signal' and `idler' in the opposite momenta. Density noise measurement in Fourier space can thus be treated as a joint detection of the $\pm \mathbf{k}$ modes (Fig.~\ref{fig:entanglement})~\cite{Chen2021ObservationInteraction} in analogy to the homodyne/heterodyne detection techniques widely adopted to characterize two-mode squeezing and quantum entanglement of biphotons in quantum optics~\cite{slusher1985observation,lvovsky2009continuous} and the spinor condensates~\cite{gross2011atomic,lucke2011twin}. 

In particular, the density noise power spectrum defined in Eq.~\eqref{eq:power_spectrum} can be expressed as the variance of the quadrature operators of quasiparticles, 
\begin{equation}
    S(\mathbf{k}) = \frac{C_k}{2}\left[ \mean{(\hat{x}_\mathbf{k}+\hat{x}_\mathbf{-k})^2} +\mean{(\hat{p}_\mathbf{k}-\hat{p}_\mathbf{-k})^2} \right] \geq C_k \,,\label{eq:sk_quadrature}
\end{equation}
where $\hat{x}_\mathbf{k}=(\hat{\alpha}_\mathbf{k}^\dagger+\hat{\alpha}_{\mathbf{k}})/\sqrt{2}$ and $\hat{p}_\mathbf{k}=i(\hat{\alpha}_\mathbf{k}^\dagger-\hat{\alpha}_{\mathbf{k}})/\sqrt{2}$ are the quadratures, and $\hat{\alpha}^{(\dagger)}_{ \mathbf{k}}$ is the quasiparticle annihilation (creation) operator, defined with the interaction $g$ at the time of measurement; $C_k = \epsilon_k/|\epsilon(k)|$ is the so-called \emph{squeezing} parameter, setting the quantum limit for the noise power spectrum. The inequality in Eq.~\eqref{eq:sk_quadrature} can be derived~\cite{Chen2021ObservationInteraction} from the Peres-Horodecki separability criterion for bipartite entanglement in continuous-variable states~\cite{duan2000inseparability,simon2000peres}. If the density matrix of the quasiparticles is separable, the inequality must be satisfied. In other words, when one measures $S(\mathbf{k})<C_k$, the quasiparticles must possess nonclassical correlations and are quantum entangled.

To make the amplified density noise squeeze below the quantum limit set by Eq.~\eqref{eq:sk_quadrature}, a second interaction quench to $g>0$ can be performed. This brings the condensate back to the superfluid regime, converting the quasiparticles to stable phonons that can accumulate real phase relative to the ground state atoms. Time evolution of the density noise power spectrum during this epoch can be explicitly expressed as $S(\mathbf{k},\tau) = C_k \left[ 1+N_\mathbf{k} + \Delta N_\mathbf{k} \cos \phi(\tau) \right]$, where $N_\mathbf{k}=\mean{\hat{\alpha}_\mathbf{k}^\dagger\hat{\alpha}_\mathbf{k}} + \mean{\hat{\alpha}_\mathbf{-k}^\dagger\hat{\alpha}_\mathbf{-k}}$ is the mean phonon number, $\Delta N_\mathbf{k}=2|\mean{\hat{\alpha}_\mathbf{k}\hat{\alpha}_\mathbf{-k}}|$ is the pair-correlation amplitude, and $\phi(\tau) = 2\epsilon(k)\tau/\hbar + \phi(0)$ is the relative phase that evolves at twice the phonon frequency. The criterion Eq.~\eqref{eq:sk_quadrature} can be violated only when $\Delta N_\mathbf{k} > N_\mathbf{k}$~\cite{robertson2017assessing,robertson2017controlling}. As time evolves, maximal two-mode squeezing for $S<C_k$ occurs when the phonon pairs are $\pi$ out of phase; maximal anti-squeezing $S>C_k$ can be observed when they become in-phase. The requirement $\Delta N_\mathbf{k} > N_\mathbf{k}$ also implies that the detected phonon pairs must be correlated and phase coherent. This suggests that quasiparticle pairs grown from MI must be seeded from quantum noise to become quantum entangled. They cannot be amplified from incoherent noise like thermal fluctuations that can diminish $\Delta N_\mathbf{k}$.

Figure~\ref{fig:entanglement} demonstrates squeezing using time-dependent density noise power spectra following the indicated interaction quench protocol: first at $g<0$ for inducing MI and then to $g>0$ to accumulate relative phonon phase $\phi(\tau)$ before imaging. The plotted power spectra $\tilde{S} =S(k)/C_k$ have been azimuthally averaged. The time axes $\tilde{\tau}=\gamma_k\tau$ in panels (a)-(c) were rescaled by the corresponding squeezing parameter $C_k$ and by the phonon frequency $\gamma_{k}=\epsilon(k)/\hbar$, respectively. Remarkably, those that had undergone MI began with large initial magnitudes $\tilde{S}>1$ and evolved towards squeezed values $\tilde{S}<1$. Panels (d)-(e) show the maximally squeezed $\tilde{S}_\mathrm{min} < 1$ and $\Delta N_{k} > N_{k}$ at large enough $k>1.5/\mu\mathrm{m}$, where initial thermal noise was suppressed and the density modulations were primarily seeded from quantum noise. The experiment carried out in Ref.~\cite{Chen2021ObservationInteraction} confirmed MI-enhanced two-mode squeezing to $\tilde{S}_\mathrm{min}\approx 0.8$ when the samples experienced a short MI period $\Delta \tau \approx 1~$ms. For longer MI period $\Delta\tau\approx2~$ms, both $N_{k}$ and $\Delta N_{k}$ increased but the measured noise showed less squeezing (larger $\tilde{S}_\mathrm{min}$), indicating that decoherence and dissipation may have quickly taken effect to reduce the coherence of quasiparticle pairs. 

\section{Summary and outlook}\label{sec:outlook}
This chapter highlights experimental studies on multidimensional attractive condensates, revealing rich dynamical nonlinear phenomena, self-trapped solitary waves, and their higher-order excitations. So far, the majority of observations have been predicted and well-described by the GPE and mean-field analyses. These demonstrations established matter waves as important testbeds for studying self-focusing nonlinear wave physics. 

There are a few interesting directions for future works. For example, a plethora of dynamical nonlinear solutions remain to be explored. Besides soliton breathers, the 1D NLSE supports generation of large amplitude rogue waves~\cite{onorato2013rogue}, Peregrine solitons, and a family of Akhmediev breathers. How these states could form spontaneously in an attractive condensate remains an open question. Can unstable multidimensional attractive condensates be further stabilized? Generally, these can be achieved through modification of the kinetic energy via a periodic potential or through modulating the interaction parameter. The former can be achieved experimentally through the application of optical lattices or DMD/SLM engineered optical potentials, while the latter can be realized through modulating the magnetic field near a magnetic Feshbach resonance (see Chapter~3.2 for more discussions). We also note that optically-induced Feshbach resonances~\cite{theis2004tuning,yan2013controlling} and the related interaction tuning techniques~\cite{bauer2009control,clark2015quantum,arunkumar2019designer} can further enable spatiotemporal control of atomic interactions. In addition, quantum fluctuations and beyond mean-field effects may also present as a stabilizing mechanism for 2D solitons, similar to quantum droplets~\cite{petrov2015quantum}.

So far, beyond mean-field effects have largely evaded experimental observations based on weakly attractive condensates. Squeezed density noise observed after a short period of MI~\cite{Chen2021ObservationInteraction} provided one rare example, suggesting that nonclassical correlations can indeed develop within an attractive condensate. If quantum entanglement could survive through the nonlinear stage of MI, \emph{fragmented} soliton-like states with macroscopic nonlocal correlations may be created. These are in line with the discussions in, for example, Refs.~\cite{kanamoto2005symmetry,mueller2006fragmentation,streltsov2008formation,weiss2009creation}. Continuing explorations of interaction quench dynamics in attractive condensates could provide test grounds for beyond mean-field theories such as the 1D Lieb-Liniger model~\cite{calabrese2007correlation,piroli2016multiparticle,yurovsky2017dissociation} and large-$N$ 2D droplets~\cite{hammer2004universal,petrov2025beyond}.

\section*{Acknowledgments}
This preprint will appear as a chapter in the Springer book entitled Short and Long Range Quantum Atomic Platforms — Theoretical and Experimental Developments (provisional title), edited by P. G. Kevrekidis, C. L. Hung, and S. I. Mistakidis.

\end{document}